\documentclass[reprint,
amsmath,amssymb,aps]{revtex4-2}
\usepackage{graphicx}
\usepackage{dcolumn}
\usepackage{bm}
\usepackage{xcolor}

\usepackage{enumitem}
\setlist{itemsep = 0.0pt}

\usepackage{booktabs}

\begin{document}

\title{Electronic structure calculations of twisted multi-layer graphene superlattices}

\author{Georgios A. Tritsaris$^a$}
\author{Stephen Carr$^b$}
\author{Ziyan Zhu$^b$}
\author{Yiqi Xie$^a$}
\author{Steven B. Torrisi$^b$}
\author{Jing Tang$^c$}
\author{Marios Mattheakis$^a$}
\author{Daniel Larson$^b$}
\author{Efthimios Kaxiras$^{a,b}$}
\affiliation{$^a$John A. Paulson School of Engineering and Applied Sciences, Harvard University, Cambridge, Massachusetts 02138, USA}
\affiliation{$^b$Department of Physics, Harvard University, Cambridge, Massachusetts 02138, USA}
\affiliation{$^c$Department of Physics, Nanjing University, Nanjing, 210093, China}

\date{\today}

\begin{abstract}
Quantum confinement endows two-dimensional (2D) layered materials with exceptional physics and novel properties compared to their bulk counterparts. Although certain two- and few-layer configurations of graphene have been realized and studied, a systematic investigation of the properties of  arbitrarily layered graphene assemblies is still lacking. We introduce theoretical concepts and methods for the processing of materials information, and as a case study, apply them to investigate the electronic structure of multi-layer graphene-based assemblies in a high-throughput fashion. We provide a critical discussion of patterns and trends in tight binding band structures and we identify specific layered assemblies using low-dispersion electronic bands as indicators of potentially interesting physics like strongly correlated behavior. A combination of data-driven models for visualization and prediction is used to intelligently explore the materials space. This work more generally aims to increase confidence in the combined use of physics-based and data-driven modeling for the systematic refinement of knowledge about 2D layered materials, with implications for the development of novel quantum devices.
\end{abstract}

\maketitle

\section{Introduction} \label{sec:introduction}
The successful isolation of graphene has motivated sustained efforts to elucidate the properties and functionality of graphene and graphene-like nanostructures \citep{Geim2013Van}. Quantum confinement endows these two-dimensional (2D) layered materials with exceptional physics and novel properties compared to their bulk counterparts. Over the years, the library of 2D materials has expanded significantly to encompass a broad range of electronic behavior from metals to insulators, with promise to induce transformational advances in applications from energy storage to quantum computing \citep{novoselov_2d_2016}. 

\begin{figure}
  \centering
  \includegraphics[width=\columnwidth]{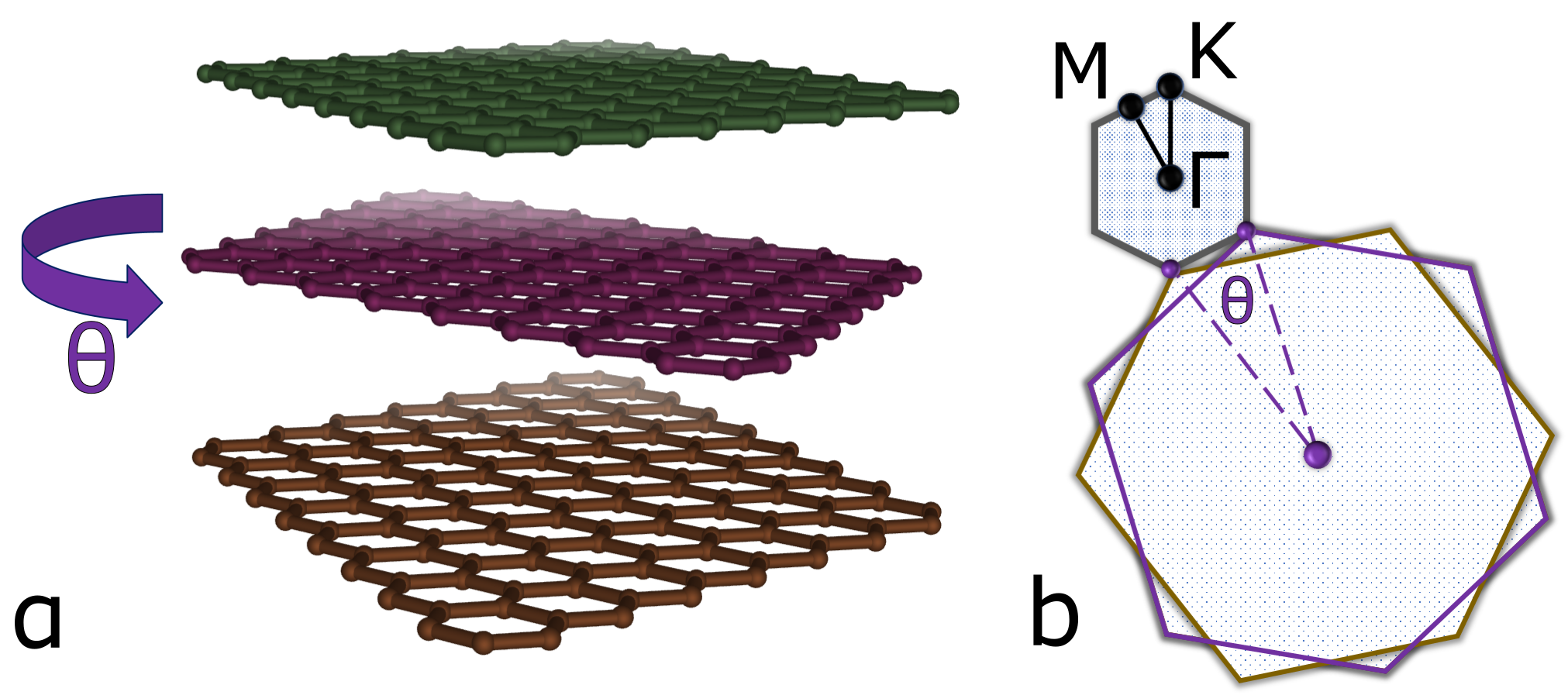}
\caption{a) Atomistic representation of a model twisted layered assembly of graphene (counterclockwise twist angle $\theta$ for any rotated layers). b) Brillouin zone of isolated layers (brown and purple outline), and twisted superlattice (grey outline) with high-symmetry points $\Gamma$, K, and M.}
\label{fig:structure}
\end{figure}

Atomically thin, single-layer forms of layered materials constitute building blocks for layered assemblies held together by weak but important interactions between the individual layers \citep{Geim2013Van,Zhou2018library,Cheon2017Data,Mounet2018Two-dimensional}. The capability to fabricate 2D architectures with specific combinations of stacking or layer orientations presents almost unlimited possibilities for devices with novel functionality emerging from the coupling of layer-specific elementary excitations. Manipulating the electronic properties of two-dimensional layered structures through their twist angle has emerged as a new paradigm in controlling the behavior of 2D layered materials and devices; this new field has been named "twistronics" \citep{Carr2017Twistronics:}. For example, 2D superlattices created by layers of graphene twisted relative to each other  (Figure~\ref{fig:structure}) have provided a platform for the study of complex electronic phenomena -- including correlated electrons and superconductivity in bilayer graphene when the two layers are twisted at the special, namely the “magic”, angle of ${\sim}1.1^\circ$  \citep{Cao2018Unconventional,Cao2018Correlated}.

There remain severe barriers that hinder the development and deployment of quantum technology based on layered graphene. An important challenge is related to the high sensitivity of electronic transport properties and current-voltage characteristics of devices on the twist angle \citep{Bistritzer2011Moire}. Although certain two- and few-layer configurations of graphene have been realized and studied \citep{Liu2019Spin-polarized,SuarezMorell2013Electronic}, the physics and properties of layered graphene assemblies remain largely unexplored: the configuration space quickly becomes enormous even with a restricted number of layers since the twist angle is a continuous parameter that critically controls measured properties. In addition, the evaluation of a novel assembly design can be resource intensive, but it remains a more tractable task compared to the inverse problem of identifying specific layered assemblies with desirable electronic behavior. Fundamental understanding of the microscopic processes that govern the electronic properties of these quantum materials is often difficult to obtain solely by electronic transport measurements. 

Materials modeling and computation have provided atomistic insights into 2D layered materials. In the prototypical case of twisted bilayer graphene, electronic band structure calculations have been used to identify electronic bands of low dispersion (or almost flat) at the Fermi level as an indicator of interesting electronic phenomena \citep{Cao2018Unconventional,Cao2018Correlated,Bistritzer2011Moire}. With regards to materials discovery, it is highly desirable to be able to efficiently search large materials spaces \citep{Tritsaris2012Methanol, Chen2016Understanding,Johannesson2002Combined,Torrisi2019,tran_active_2018}. High-throughput (HT) techniques are particularly useful for assessing the scope and limitations of novel materials concepts and discovering new materials, and various approaches have been developed tailored for 2D layered materials. For instance, Mounet {\em et al.} \citep{Mounet2018Two-dimensional} relied on HT calculations to assess the ease of exfoliation of single-layer forms of 2D layered materials from their experimentally known bulk counterparts; Bassman {\em et al.} \citep{Bassman2018Active} combined a Gaussian regression model with density functional theory (DFT) calculations to identify layered assemblies with target band gap; the work of Haastrup {\em al.} \citep{Haastrup2018Computational} used HT {\em ab initio} calculations based on DFT and many-body perturbation theory to establish a 2D materials database for the computational modeling and design of new 2D layered materials. This modality of materials discovery for graphene-based layered assemblies is now possible owing to the continuing development of quantum mechanical methods for electronic structure calculations of layered graphene  \citep{Cao2018Unconventional, Cao2018Correlated, Tritsaris2016Perturbation, Fang2016Electronic}. In contrast, a HT approach for the study and discovery of layered assemblies remains impractical in the lab due to the time required to fabricate and characterize a device, with typical rates of production being only few experiments on a new material in a month or year. 

A HT computational approach necessitates dedicated tools to efficiently and seamlessly execute, manage, and visualize hundreds or thousands of calculations. A sufficient number of computational frameworks exist for HT materials calculation \citep{Jain2015FireWorks:,Curtarolo2012AFLOW:,Pizzi2016AiiDA:}, although they tend to rely on extensions to take into consideration some important aspects of the materials knowledge creation cycle such as optimization and decision-making for materials selection and design. In the paradigm of materials informatics, data-driven approaches are used to extract broadly applicable physical insights from large materials data sets by identifying minimal sets of structural and functional descriptors. They are also used to expand the design space and produce informative visualizations of materials spaces for the accelerated discovery of not-yet-developed materials systems \citep{Sun2019Accelerated,Bassman2018Active,Botu2015Adaptive,Cubuk2015Identifying,Janet2019quantitative,Sparks2016Data}.

Here we introduce theoretical concepts and methods for the creation, combination, and use of materials information and apply them to specifically investigate the electronic structure of graphene-based layered assemblies. To that end, we use workflows that combine physics-based (tight binding; TB) and data-driven (e.g. machine learning) models to obtain insights into the electronic structure of layered graphene and intelligently search the space of multi-layer assemblies for potentially interesting electronic behavior. 

The manuscript is organized as follows: Section~\ref{sec:concepts} introduces concepts, models and methods for the creation, combination, and use of materials information. Algorithmic, computational, and implementation details are provided therein. Section~\ref{sec:results} provides a critical discussion of trends in the TB band structures of multi-layer assemblies of graphene, obtained and analyzed using the methodology introduced in Section~\ref{sec:concepts}. We also attempt to identify specific layered assemblies that might exhibit unusual electronic properties. Finally, Section~\ref{sec:conclusions} summarizes findings and proposes future research directions.

\section{Concepts, models and methods} \label{sec:concepts}
\begin{figure}
  \centering
  \includegraphics[width=\columnwidth]{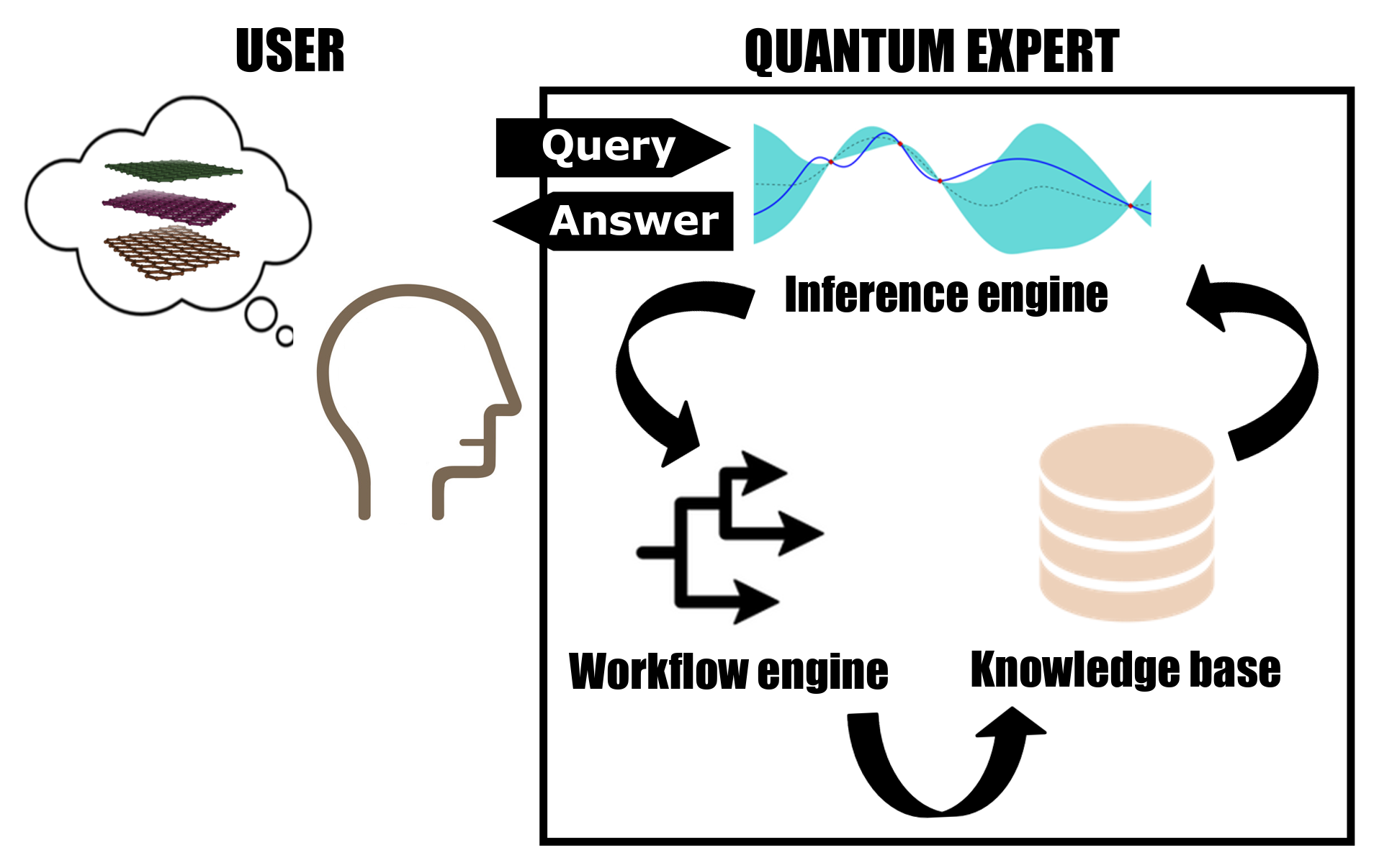}
\caption{Schematic and modus operandi of an {\em in silico Quantum Expert} for the creation, combination, and use of materials information. Key components are an inference engine for predictions, a workflow engine for calculations, and a database system.}
\label{fig:expertsystem}
\end{figure}

The creation of materials knowledge begins with the introduction of a materials concept to be subsequently evaluated. The findings are then codified and combined with existing materials knowledge to generate novel concepts, completing a cycle. At the confluence of artificial intelligence and knowledge management, knowledge-based systems such as decision support systems and expert systems constitute a technology that seeks to facilitate the accessibility and dissemination of information within a particular knowledge domain \citep{Shu-HsienLiao2005Expert, Turban2004Decision}. We propose the unifying conceptual framework of an {\em in silico Quantum Expert} (Figure~\ref{fig:expertsystem}) for the processing of materials information, here implemented as a system specifically for the study and design of 2D superlattices of graphene.
 
The key components are:
\begin{itemize}
    \item an inference engine for materials predictions, 
    \item a workflow engine for quantum mechanical calculations, 
    \item and a database system for storing materials data and information. 
\end{itemize}
Figure~\ref{fig:expertsystem} presents a schematic of the modus operandi:
\begin{enumerate}
    \item The user queries the system for information about one or more materials. For our particular application, the queries correspond to questions of the form: "{\em what is the likelihood of low-dispersion electronic bands near the Fermi level in the single-particle band structure of a specified layered assembly of graphene?}".
	\item The system responds with the requested information, which in our test case is a prediction for intresting electronic properties. These predictions require prior calculation and analysis of the low-energy band structures of layered assemblies for which an atomistic representation based on commensurate supercells exists (arbitrary rotations generally involve incommensurate, i.e., not periodic, 2D lattices even for layers of the same material) \citep{Carr2018Relaxation}.
	\item If these calculations have not been already performed, the system proceeds to execute a set of predefined workflows for quantum mechanical calculations and add the results to the knowledge base. In this application, the calculations of layered assemblies consist of setting up and solving an {\em ab initio} TB model.
\end{enumerate}
We provide application-specific implementation details for each component below. 

{\em Inference engine}. The inference engine uses available materials information to produce new information. Multiple data-driven models are used to obtain a prediction for interesting electronic properties for an input twisted layered assembly. Each model, $m = (e, h)$, is defined by an estimator, $e$, and a set of hyperparameters, $h$, and relies on a measure of band flatness in calculated TB band structures of (approximate) commensurate supercells to provide predictions for arbitrary twist angle $\theta$ by means of quadratic interpolation. As we explain in more detail below, one estimator we employed relies on the Hough transform and another on the standard deviation of band energy eigenvalues. For each model $m$, a prediction $p_m(\theta_m^*)=1$ is assigned to all twisted layered assemblies that maximize band flatness, and the prediction 
$$p_m(\theta)=\rm{exp}(-\frac{|\theta_m^*-\theta|}{0.1^\circ})$$
is assigned to all twist angles near a magic angle $\theta_m^*$. The denominator 0.1$^\circ$ is fixed and chosen to reflect the sensitivity of calculations and measurements with respect to the twist angle. For each of the two estimators $e$, the average value, $p_e(\theta)$, over the corresponding $h$ is calculated, and the prediction 
$$p(\theta) = \rm{max}_e(p_e(\theta))$$ 
is assigned to the input layered assembly, in a blend that purposefully favors false positives. In the limiting case $p_e(\theta) = \{0, 1\}$ such blending produces the same results as the logical OR operation -- another reasonable choice would have been $p(\theta)=1-\prod_e(1-p_e(\theta))$. The absolute difference between the two averages $p_e(\theta)$ is used to quantify uncertainty across estimators in terms of precision.

{\em Workflow engine}. The function of this component is to augment the knowledge base using quantum mechanical calculations. In this work we elect to use FireWorks (version 1.9.2) as the workflow engine, a Python-based library for defining, managing, and executing workflows in a decentralized fashion on different types of high-performance computing resources \citep{Jain2015FireWorks:}. 

\begin{figure*}
  \centering
  \includegraphics[width=\textwidth]{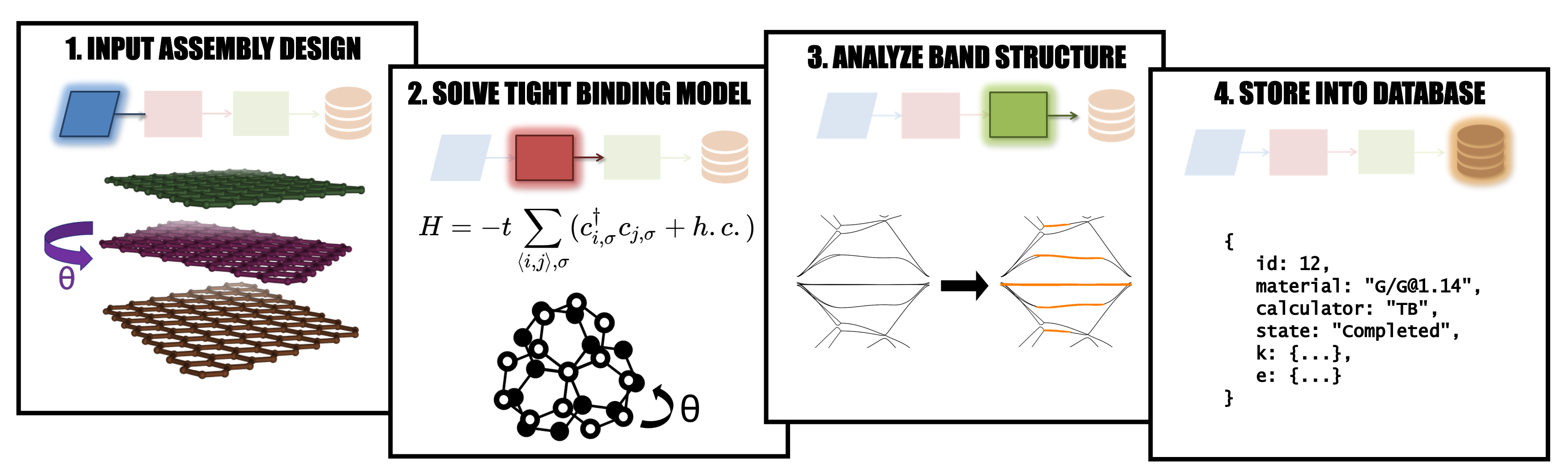}
\caption{Schematic of high-throughput workflow for physics-based calculation and data-driven analysis of the electronic band structure of 2D layered assemblies.}
\label{fig:workflow}
\end{figure*}

Materials calculations that are based on a quantum mechanical description of electronic structure can be computationally demanding, but DFT favorably balances computational cost with accuracy \citep{kohn_self-consistent_1965}. Even for superlattices of the same material however a twist can render atomic-scale modeling and calculation practically intractable due to the size of the resulting atomistic models. For example, commensurate supercells of bilayer graphene with twist angles near the magic angle comprise more than 10,000 atoms. We use electronic structure calculations based on an TB model using parameters extracted from simple and accurate DFT calculations. The corresponding computational workflow entails four main tasks (see also Figure~\ref{fig:workflow}): 
\begin{enumerate}
    \item Parse a string representing an input layered assembly into a list of rotations that is used to specify an atomistic model and the corresponding TB Hamiltonian.
    \item Generate an effective TB model for in-plane (8 nearest neighbors) and inter-plane p$_z$ orbital interactions, without any adjustable parameters, using a basis of maximally localized Wannier functions. The electronic levels are then calculated by efficient diagonalization of the system’s Hamiltonian. The memory required scales as $\mathcal{O}(N^2)$, where $N$ is the number of atoms in the supercell. A detailed presentation of the method is provided in the works of Fang {\em et al.} \citep{Fang2016Electronic,Fang2015Ab} Here we use 60 points to sample the Brillouin zone along the direction connecting the high-symmetry k-points $\Gamma$, M, and K. 
	\item Analyze the calculated energy levels in a window of 0.30~eV centered at the Fermi level using two different approaches to the detection of (almost-)flat bands in the TB band structure. The first approach relies on an image-based Hough line transform, as implemented in OpenCV (version 4.1.1), an open source computer vision and machine learning software library \citep{Bradski2000OpenCV}. The Hough line transform converts the problem of detecting collinear points into the problem of finding concurrent curves. The sensitivity of the algorithm depends on a so-called accumulator threshold, here between 50 and 200 (the smaller this number is, the more linear segments are detected, maybe incorrectly). The second approach relies on a calculation of the contribution of each k-point to the standard deviation of energy eigenvalues for each band. A band is classified as flat if the standard deviation is lower than a threshold value, here between 5~meV and 15~meV, after removing between 20\% and 50\% of k-points with the largest contribution to the standard deviation. The outputs of the two methods are complementary since the first circumvents the issue of artificial discontinuities in the electronic bands but the latter considers information about overlapping bands that is necessarily lost in a graphical representation although important near degeneracies. 
	\item Insert the electronic band structure and post-processing meta-data  into the database for later access.
\end{enumerate}
In summary, the above computational workflow transforms a string representing a layered assembly into measures of band flatness by means of efficient diagonalization of a TB Hamiltonian and featurization of the calculated band structures.
 
{\em Database system}. This component serves as the storehouse of materials data and information (the knowledge base). In this work the database relies on a document-based data model, as implemented in the document-oriented (NoSQL) database MongoDB. Unlike relational databases, MongoDB is easy to scale while allowing the content and size of documents to differ from one to another. To reduce maintenance burden, the same database is used to support workflow management as required by Fireworks (e.g., workflow state and dependencies), and store the inputs and outputs of materials calculations, including structural details of the input layered assembly (e.g., rotations), its calculated TB band structure (i.e., a list of k-point dependent energy eigenvalues), and accompanying meta-data. 

We make use of a minimal programmatic interface as a natural interface to the Python-based libraries used to implement the components. For the sake of simplicity, the query is restricted to strings representing twisted layered assemblies and the response is restricted to the corresponding predictions $p(\theta)$. In a Bayesian framework, the predictions $p(\theta)$ constitute prior probabilities to be updated after more careful examination.

Overall, our computational strategy is based on scalable, parallelizable, decentralized materials calculations and it is distinctly higher up on the ladder of abstraction compared to other frameworks for computational materials science \citep{Jain2015FireWorks:, Curtarolo2012AFLOW:, Pizzi2016AiiDA:}. We note in passing that although the conceptual framework of Quantum Expert is useful for organizing materials calculations and directing the collection and analysis of data, it is not necessary for reproducing our specific findings. We also purposefully use only established approaches and open source libraries.

Coupling and linking the various components but also individual tasks within workflows pose challenges like devising systems and protocols for efficient codification and exchange of materials information. The first step in this direction is to be able to name a specific layered assembly in an unambiguous fashion. Despite the maturity of chemical language and notation such as SMILES and SMARTS for representing chemical compounds \citep{Weininger1988SMILES, D.C.I.SystemsSMARTS} these cannot describe critical features of a layered assembly such as the twist angle between two neighboring layers. Previously, we introduced a flexible layered assemblies notation (LAN) \citep{Tritsaris2019LAN}, which we use to enumerate structures in the database, unambiguously describe trends across layered assemblies of arbitrary relative orientations of the constituting graphene layers, and construct structural descriptors for machine learning. The notation derives from a theoretical materials concept of layer-by-layer assembly of layered structures using a sequence of rotation, vertical stacking, and other operations on individual 2D layers. For example, the string ‘G/G@1.08’ describes a bilayer of graphene (often referred to in the literature as ‘TBG’) with twist angle $\theta = 1.08^\circ$, the string ‘(G/G)/(G/G)@1.23’ (or the shorter ‘2G/2G@1.23’) describes a twisted double bilayer of graphene (‘TDBG’ or ‘TBBG’) with $\theta = 1.23^\circ$, and so on. The symbols ‘/’ and ‘@’ are used to describe the binary operations of the vertical stacking of a layer or layered (sub)structure on another, and counterclockwise rotation of a layer or layered (sub)structure by some angle about the stacking direction (in degrees; $360^\circ = 2\pi$), respectively. The definition of the underlying grammar and basic operations is provided in Tritsaris {\em et al.} \citep{Tritsaris2019LAN}. 

\section{Results and Discussion} \label{sec:results}
We proceed to apply our concepts, models and methods to investigate the electronic structure of twisted assemblies of graphene with two or more layers.

\subsection{The simplest assembly}
We first concentrate attention on the prototypical system of twisted bilayer graphene, G/G@$\theta$. This configuration has been commonly used as a platform for the study of non-trivial emergent physical behavior that can be controlled with great precision with the relative rotation between the two layers. When the two layers are twisted at the magic angle of $1.1^\circ$ it is possible to reduce electrical resistance significantly \citep{Cao2018Unconventional,Cao2018Correlated}. Although elucidating the nature of superconductivity in graphene remains the goal of much ongoing research, features such as low-dispersion electronic bands in the single-particle band structure have been established as indicators of such unconventional physics \citep{Bistritzer2011Moire}.

\begin{figure*}
  \centering
  \includegraphics[width=\textwidth]{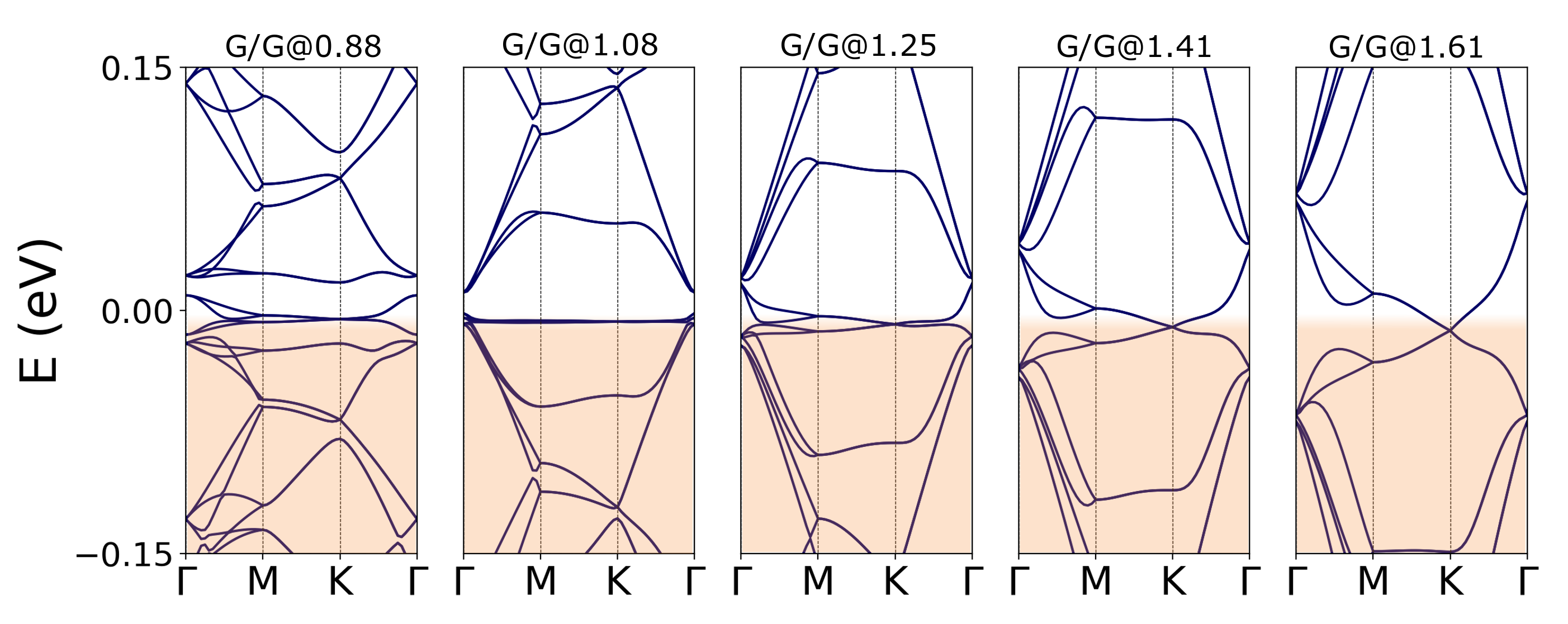}
\caption{Low-energy electronic band structure for twisted bilayers of graphene near the “magic” angle of $1.1^\circ$, obtained using an {\em ab initio} tight binding model. The shaded area marks occupied electronic states.}
\label{fig:blg}
\end{figure*}

We begin by requesting predictions $p(\theta)$ for $0.88^\circ \leq \theta \leq 21.79^\circ$. For twist angles corresponding to commensurate 2D superlattices, the band structure is calculated using TB on the basis of which predictions are obtained for the entire range of twist angles. The twisted superlattices consist of alternating AA- and AB-stacked regions, while the Brillouin zone folds as shown in Figure~\ref{fig:structure}b. Each atomistic model of commensurate superlattices consists of two graphene layers with in-plane lattice constant of 2.47~$\rm{\AA}$, while the distance between the two graphene layers is fixed at 3.35~$\rm{\AA}$. Twist angles that result into commensurate stackings are identified following the formalism of Uchida {\em et al.} \citep{Uchida2014Atomic}, 
which associates an angle $\theta$ with a pair of integers $(M, N)$ that define the periodicity of the layers as:
$$cos(\theta)=\frac{N^2+4NM+M^2}{2(N^2+NM+M^2)}.$$
For instance, in the regime of small twist angles, the three pairs $(M, N)$ of (32,31), (27,26), and (23,22), correspond to $\theta$ of $1.05^\circ$, $1.25^\circ$, and $1.47^\circ$, respectively. In total, 33 unique commensurate bilayers for $0.88^\circ \leq \theta \leq 21.79^\circ$ are calculated (21 for 1.05$^\circ \leq \theta \leq 2.88^\circ$) and analyzed. 
For large $\theta$ the low-energy band structure resembles that of an isolated graphene layer. As $\theta$ becomes smaller, the number of bands increases and the electronic bands near the Fermi level become flatter. Figure~\ref{fig:blg} shows TB band structures of bilayers with $\theta$ near the magic angle. In this range, strong interlayer hybridization leads to the formation of low-dispersion bands. For even smaller rotations ($\theta < 1.0^\circ$) structural relaxation effects have been shown to modify the band structure significantly \citep{Nam2017Lattice,Carr2018Relaxation,Carr2019Exact}. These effects however are not accounted for in the current computational workflow to reduce computational burden and for that reasons small twist angles are not discussed here. 

An important observation is that the electronic bands evolve smoothly in the regime of small rotations (Figure~\ref{fig:blg}), which allows for predictions for incommensurate superlattices using interpolation as described in Section II. As it would be expected from previous theoretical and experimental results, $p(\theta)$ is maximized for $\theta^* = 1.1^\circ$ (see G/G@1.08 in Figure~\ref{fig:blg}) \citep{Cao2018Unconventional,Cao2018Correlated,Bistritzer2011Moire}.

\subsection{Three or four layers of graphene}
We proceed to use our computational strategy to investigate twisted layered assemblies of graphene with three or four AB-stacked layers. We limit our investigation to configurations with a single twist angle for any rotated layers resulting in two and five configurations for an input angle for the tree- and four-layer assemblies, respectively. The three-layer configurations correspond to the assemblies of a twisted monolayer on a bilayer of graphene, 2G/G@$\theta$, and a twisted graphene monolayer encapsulated by another two, G/G@$\theta$/G. The four-layer configurations include assemblies such as a twisted monolayer on a trilayer, 3G/G@$\theta$, and a twisted bilayer on a bilayer, 2G/2G@$\theta$. 

\begin{figure*}
  \centering
  \includegraphics[width=\textwidth]{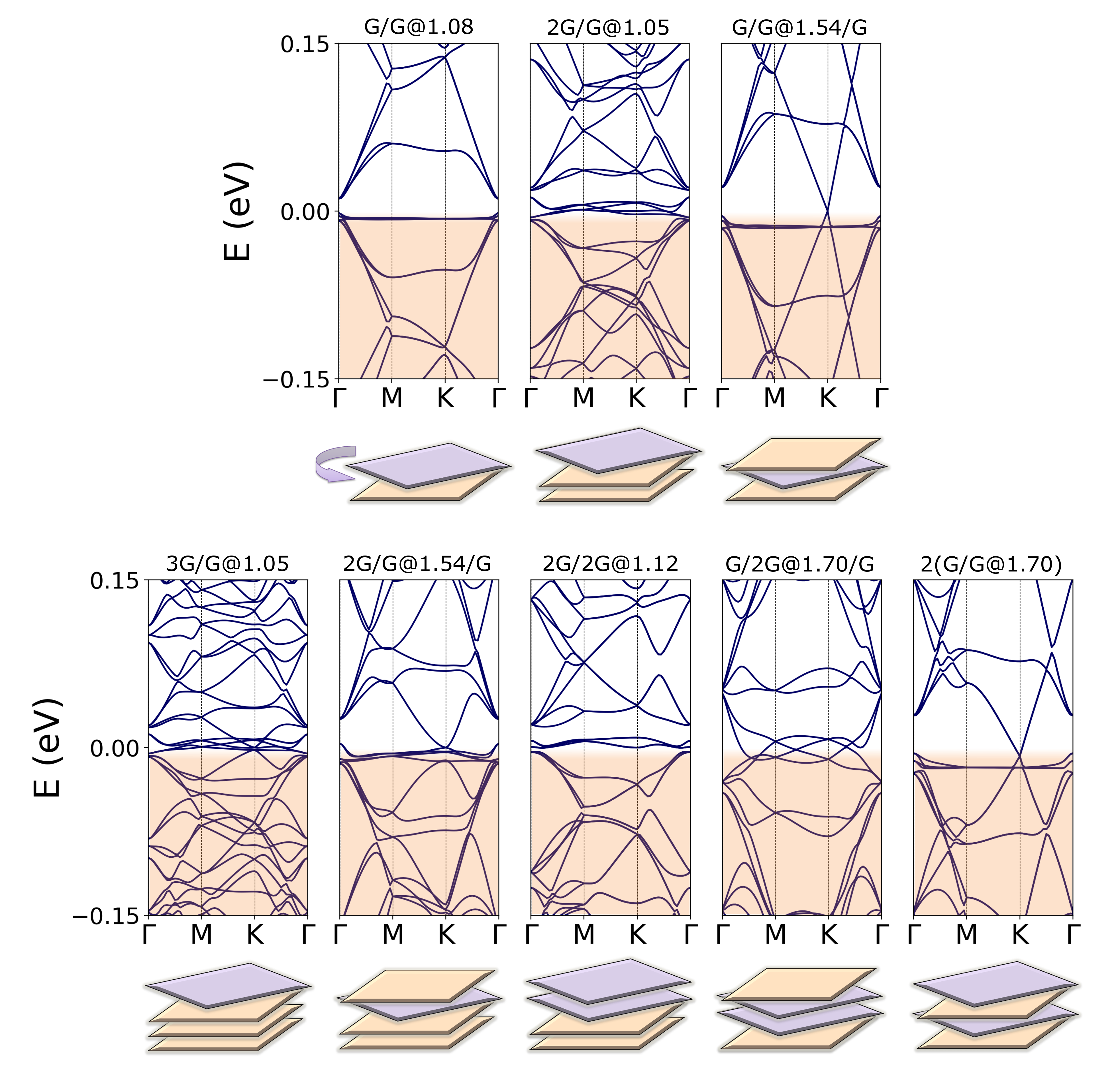}
\caption{Low-energy electronic band structure for twisted three- and four-layer assemblies of graphene with low-dispersion or almost-flat electronic bands, obtained using an {\em ab initio} tight binding model. The shaded area marks occupied electronic states.}
\label{fig:34L}
\end{figure*}

We obtain predictions $p(\theta)$ for $0.88^\circ \leq \theta \leq 21.79^\circ$, which requires calculation and analysis of a total of 231 additional TB band structures of commensurate 2D superlattices. Using $p(\theta)$ as a guide, we inspect the band structures of commensurate supercells near $\theta^*$ and verify low-dispersion bands (Figure~\ref{fig:34L}). Some of the important features in the TB electronic band structures are discussed below:

{\em 2G/G@$\theta$}. The electronic structure of a twisted monolayer on a AB-stacked bilayer of graphene resembles that of the twisted bilayer with $\theta^* = 1.1^\circ$. The reduced symmetry causes the electronic bands near the Fermi level to be more dispersive than the twisted bilayer.

{\em G/G@$\theta$/G}. This sandwiched configuration achieves bands with very low dispersion near $\theta^* = 1.5^\circ$. A distinguishing feature of this configuration’s band structure is a Dirac cone that resembles that of an isolated graphene layer. The Dirac cone feature is caused by an anti-symmetric combination of electronic orbitals (the top and bottom layers are in registry and couple identically to the middle layer), and is not sensitive to the twist angle \citep{Carr2019Coexistence}. These findings are in excellent agreement with the work of Khalaf {\em et al.} \citep{Khalaf2019Magic}, which used model Hamiltonians to argue that this trilayer configuration’s magic angle is obtained by multiplying the bilayer magic angle by $\sqrt{2}$.

{\em 3G/G@$\theta$}. The twisted monolayer on a trilayer of graphene has $\theta^* = 1.1^\circ$. Unlike the twisted monolayer on bilayer, the flat bands are not gapped from the rest of the spectrum. A cone-like feature has begun to emerge at the K-point. As the number of layers in the unrotated part of the stack increases, the low-dispersion bands of the 2D twisted interface will coexist within a continuum of bands from bulk graphite \citep{Cea2019Twists}.

{\em 2G/G@$\theta$/G}. A twisted monolayer of graphene, when twisted on an bilayer of graphene, has a magic angle of $\theta^* = 1.5^\circ$.  Like the case of the sandwiched trilayer of graphene, the increase in the magic angle is caused by the enhanced interlayer coupling, but now parabolic bands meet near the Fermi level instead of a Dirac cone. This is comparable to the change in the electronic structure when going from the twisted bilayer to the twisted monolayer on bilayer, as the substitution of the bottom layer with a Bernal-stacked bilayer has reduced the symmetry and increased the dispersion of the low energy bands.

{\em 2G/2G@$\theta$}. Twisted double bilayer graphene has been approached as an alternative to bilayer graphene in exploring correlated insulators and superconductivity \citep{Lee2019Theory}. Electrically tunable half-filled Mott-like insulating states for a wide range of twist angles, and superconductivity with critical temperature onset at 12 K, has been demonstrated in devices of this configuration \citep{Shen2019Observation,Cao2019Electric}. We identify a region of almost-flat bands at the Fermi level near $\theta^* = 1.1^\circ$, although their dispersion is larger compared to bilayer graphene due to the absence of C2 symmetry.

{\em G/2G@$\theta$/G}. Electronic bands with low dispersion are identified around $\theta^* = 1.6^\circ$ in the case of a sandwiched bilayer of graphene. This generalization of the sandwiched monolayer lacks the symmetry-protected Dirac cone feature at the K-point, and the resulting electronic structure more closely resembles that of the twisted monolayer on bilayer configuration. The pairs of effective Dirac cones at both the M-point and K-point may warrant further study.

{\em 2(G/G@$\theta$)}. This configuration belongs to the same class of layered assemblies of graphene with alternating relative twist angle as the sandwiched trilayer. It also exhibits magic angle flat bands, coexisting with a pair of Dirac bands at the moir{\'e} K-point. For this specific configuration, according to Khalaf {\em et al.} \citep{Khalaf2019Magic} the magic angle is expected to be $\sim$1.62 times larger than this of bilayer graphene ($\theta^*$ = 1.8$^\circ$). Very flat electronic bands are indeed identified at $\theta^* = 1.7^\circ$. 
 
\begin{figure*}
  \centering
  \includegraphics[width=\textwidth]{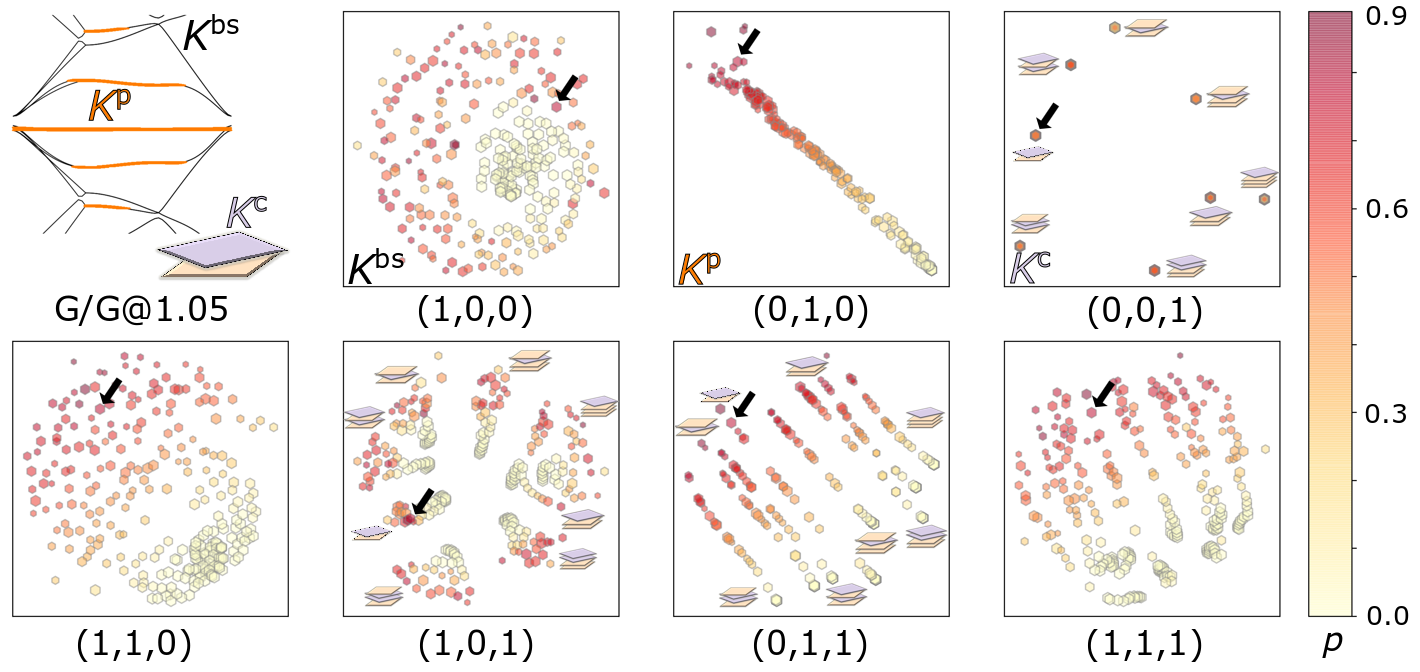}
\caption{Visualizations of the materials space using multidimensional scaling (MDS) with different combinations $(a, b, c)$ of three dissimilarity matrices $K\rm{^{bs}}$ (based on the calculated band energies), $K\rm{^{p}}$ (based on the probabilities $p(\theta)$), and $K\rm{^{c}}$ (based on configurations). Points are color-coded with respect to predictions $p(\theta)$ for low-dispersion electronic bands in the band structure (white for small; red for high) and the symbol size is proportional to the associated uncertainty (larger for higher precision). The black arrows indicate where the same layered assembly, G/G@1.05, can be found in each panel. The top left panel shows its band structure, with flat segments marked in orange (Hough transform).}
\label{fig:atlas}
\end{figure*}

As the number of calculated materials increases it becomes impractical to inspect calculations one at a time. To remedy the situation, we generate and inspect information-rich representations of the database in a state that contains all calculations of commensurate supercells (Figure~\ref{fig:atlas}). To produce these visualizations we rely on data embedding using multidimensional scaling (MDS), as implemented in the scikit-learn package (version 0.20.3), a machine learning library for Python \citep{Pedregosa2011Scikit-learn:}. MDS as an approach to dimensionality reduction is particularly useful for visualizing the level of (dis)similarity of objects. Here, these objects are the TB band structures, predictions $p(\theta)$, and vectorized representations of the layered assemblies (264 in total). We use MDS to translate them into sets of 2D points in abstract Cartesian spaces by way of a  264-by-264 dissimilarity matrix $K = [k_{ij}]$ with elements $k_{ij}$ defined as:
$$k_{ij}=||(a k_{ij}^{\rm{bs}}, b k_{ij}^{\rm{p}}, c k_{ij}^{\rm{c}})||_2,$$ 
with $a, b, c = \{0, 1\}$. Its calculation relies on combinations $(a, b, c)$ of three different metrics for evaluating pairwise distances between objects ($k_{ii} = 0$):
\begin{itemize}
	\item The calculation of $K\rm{^{bs}}$ relies on Euclidean distances between the TB band structures in an energy window of 0.30~eV centered at the Fermi level. The band structures are represented as binary images transformed by means of singular value decomposition to 200-dimensional feature vectors (covering 97\% of variance).
	\item The calculation of $K\rm{^p}$ relies on absolute differences between the predictions $p(\theta)$.
	\item The calculation of $K\rm{^c}$ relies on cosine distances between unit-norm vectorized representations of the layered assemblies. In analogy to a “sum over bonds” vector for the featurization of chemical compound spaces \citep{Elton2018Applying}, we construct a {\em sum over interfaces} (SOI) representation. These are vectors composed of sums, where each sum represents a counting of a particular interface type (G/G, G/G@$\theta$, G/G/G, G/G/G@$\theta$, etc.). Two- and three-layer groupings (12 interfaces in total) suffice to uniquely describe the space of two- to five-layer assemblies, a reasonable choice given the weak but still important interactions between neighboring layers. By standard definition, cosine (dis)similarity metrics reduce the importance of the size of the assemblies.
\end{itemize} 

Three visualizations are shown in Figure~\ref{fig:atlas}, one for each of the above three matrices, which we use as building blocks for another four matrices corresponding to combinations $(a, b, c)$ of (1,1,0), (1,0,1), (0,1,1), and (1,1,1). In Figure~\ref{fig:atlas}, the points are color-coded with respect to $p(\theta)$ and the symbol size is proportional to the associated uncertainty (larger for higher precision with respect to $p_e(\theta)$). The black arrows in Figure~\ref{fig:atlas} indicate where the same layered assembly, G/G@1.05, can be found in each panel. As is often the case with visualizing materials information, the elected estimator MDS or approach to constructing the dissimilarity matrix K is neither unique nor necessarily optimal. Analogous representations are obtained by using the t-distributed stochastic neighbor embedding (t-SNE) method. The generated 2D plots together constitute {\em an atlas of interface complexity} that upon inspection reveals distinct groupings. With regards to materials discovery, such data-driven representation of the materials space simplifies the task of screening for interesting electronic behavior, drastically narrowing the list of potential candidates for more careful consideration. 

These representations provide a clear picture of relationships across the entire materials database that is not readily observable by means of enumeration: 
\begin{itemize}
    \item $K\rm{^{bs}}$ favors a clustering with respect to electronic bands, 
    \item $K\rm{^p}$ favors a one-dimensional ranking with respect to the predictions $p(\theta)$, and 
    \item $K\rm{^c}$ favors a clear clustering with respect to configurations (type of interfaces). 
\end{itemize}
Combinations of the three matrices combine materials information to highlight multiple trends simultaneously: for example, a combination of (0,1,1) separates configurations while sorting structures according to $p(\theta)$, which can be interpreted as a score for selecting potentially interesting structures for closer inspection. Interestingly, a combination of (1,0,1) generates a sorting roughly corresponding to $p(\theta)$ even if information from $K\rm{^p}$ is not used. Only very few predictions are associated with relatively high uncertainty. For example, one case is associated with multiple artificial band crossings at very low twist angles for 3G/G@$\theta$. For the purpose of materials discovery, after rounding $p(\theta)$, the task reduces to transforming a list of rotations (0 or $\theta$) into a yes/no response for more careful examination.

\subsection{Other layered assemblies}
We next extend our exploration to systems with more than four layers. Within the context of inverse materials design, our implementation of the Quantum Expert can also be used to answer questions of the form: "{\em what is the structure of a novel 2D layered assembly that could exhibit interesting electronic properties?}" 

A pedestrian approach involves the use of the 2D embedding (0,1,1) shown in Figure~\ref{fig:atlas}, i.e., using a combination of the dissimilarity matrices $K\rm{^p}$ and $K\rm{^c}$. Informed by our findings for the three- and four-layer assemblies with alternating relative twist angle, we use it to identify $\theta^*$ that maximize $p(\theta)$ for the five-layer configuration 2(G/G@$\theta$)/G. First, we vectorize the structures according to SOI and then seek the nearest neighbors in the 2D embedding. With respect to cosine distances the two most similar configurations are G/G@$\theta$/G and G/G@$\theta$/G/G@$\theta$, while 3G/G@$\theta$ is the least. The nearest neighbors are associated with twist angles around $1.1^\circ$ and $1.6^\circ$. 

\begin{figure*}
  \centering
  \includegraphics[width=\textwidth]{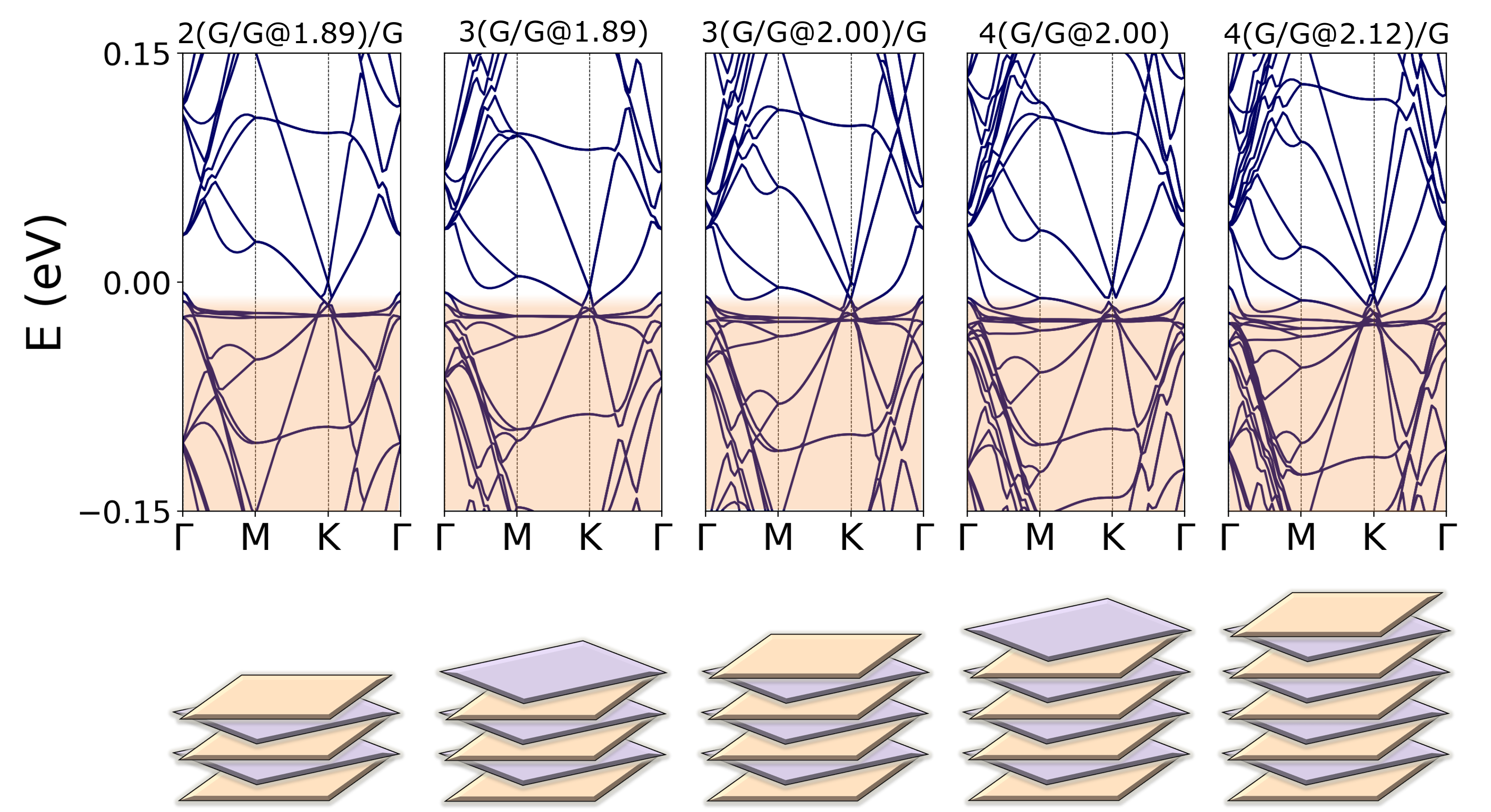}
\caption{Low-energy electronic band structure for alternating-twist layered assemblies with five to nine graphene layers with flat electronic bands, obtained using an {\em ab initio} tight binding model. The shaded area marks occupied electronic states.}
\label{fig:5L}
\end{figure*}

Another approach would rely on the use of a separate estimator (or optimizer) in the role of the researcher requesting and using the predictions $p(\theta)$ for black-box optimization and adaptive search of the materials space \citep{roch_chemos:_2018,Dunn_2019}. We demonstrate this modality of materials discovery by employing an extremely randomized trees (a.k.a. extra-trees) model for regression \citep{Geurts2006Extremely}, as implemented in the scikit-learn machine learning library \citep{Pedregosa2011Scikit-learn:}. Picking interesting candidate structures for detailed investigation implies decision. Extra trees is an ensemble-based estimator that fits a number of randomized decision trees and predicts through averaging the predictions of all trees to control over-fitting. We use SOI vectors augmented with twist angles to train a forest of 100 trees of maximum depth of 8 for regression. These hyperparameters were obtained using 30-fold cross-validation (average coefficient of determination $R^2$ of 0.7). By computing feature importances, we verify the twist angle to be by far the most important and therefore strongly correlated with $p(\theta)$, followed by the number of G/G@$\theta$ (or G@$\theta$/G) and G/G@$\theta$/G interfaces. The trained meta-model codifies a structure-property relationship leading to results in good agreement with full TB calculations, $\theta^* = 1.9^\circ$ (error of 15\%; see also Figure~\ref{fig:5L}). 

These findings motivate us to look at general trends in the electronic structure of alternating-twist layered assemblies. According to the work of Khalaf {\em et al.} \citep{Khalaf2019Magic}, for alternating-twist layered assemblies the magic angle is expected to reach the limiting value of $2.2^\circ$ as the number of layers increases. To test this theoretical prediction, but also to demonstrate the capabilities of our models and methods, we obtain $p(\theta)$ for layered assemblies with up to twenty layers of graphene with twist angle around $2.0^\circ$. We find that the nine- and twenty-layer assembly exhibit overlapping flat bands in the corresponding TB band structure at $\theta^* = 2.1^\circ$ (Figure~\ref{fig:5L}). We ascribe this relatively small discrepancy (5\%) to the higher accuracy of our {\em ab initio} TB approach compared to the model Hamiltonians used in the aforementioned study. The generally good agreement between theoretical predictions, full TB calculations, and data-driven regression in retrospect validates SOI for representing structures.

In summary, the materials knowledge creation cycle can be outlined as follows:
\begin{enumerate}
    \item Define the objective as the identification of twisted layered assemblies of graphene with potentially interesting electronic behavior.
	\item Calculate and analyze the band structures of a set of layered assemblies and use them to predict such behavior.
	\item Train an estimator for predictions of layered assemblies not contained in the database.
	\item Identify promising candidate structures from step 3 and return to step 2 to assess them by full physics-based calculations. 
\end{enumerate}
In other words, the predictions $p(\theta)$ generated in step 2 are used to support decision making in what can be thought as a basic active learning approach for autonomous (closed-loop) exploration of the materials space. Alternatively, the extra-trees regressor can be used as another predictive data-driven model available directly to the inference engine for faster and broader predictions that become increasingly more accurate as the knowledge base expands. Conducting comprehensive model selection or an extensive exploration of the space of twisted layered assemblies with five or more graphene sheets nevertheless remains beyond the scope of this work.

\section{Conclusions} \label{sec:conclusions}
We introduced the unifying conceptual framework of an automated {\em in silico Quantum Expert} for the creation, combination, and use of materials information. This part is completely general and should also be applicable to the investigation, optimization and screening of materials other than those studied here, describing in a uniform way such workflows \citep{Bassman2018Active,tran_active_2018}. We implemented it specifically to identify graphene-based assemblies for interesting electronic behavior as an application. Materials calculations relied on an {\em ab initio} tight binding model. Their analysis was based on various featurizations, including an approach based on computer vision (Hough transform) to construct measures of band flatness, and {\em sum over interfaces} descriptors for the studied structures. We further produced informative visualizations of the materials space, or {\em an atlas of interface complexity}, by means of embeddings in two dimensions. The various transformations of materials information were enabled by a flexible {\em layered assemblies notation}, which we used to enumerate structures in the database so that we could unambiguously describe trends across configurations. 

Structural relaxation effects have been shown to modify the band structure significantly in the regime of small twist angles in twisted bilayer graphene \citep{Carr2019Coexistence}. These are not accounted for in the current computational workflow, but can be important even for configurations beyond the twisted bilayer. For instance, using a continuum model for interlayer interactions, Zhu {\em et al.} \citep{Zhu2019Moire} showed that the relaxation patterns of twisted trilayer of graphene are “moir{\'e} of moir{\'e}” as a result of the incommensurate coupling of two bilayer moir{\'e} patterns. An obvious extension of the current work therefore would include atomic relaxation. In addition, external conditions and perturbations such as pressure, temperature or a magnetic field could be considered as they are used to control the properties of a layered assembly. Here, we considered only high-symmetry directions in the Brillouin zone for calculation and analysis of electronic band structures, but more detailed physical insights could be obtained by densely sampling the entire Brillouin zone. The use of more refined methods for physics-based calculations, for example using more comprehensive model Hamiltonians and data-driven analysis with more extensive model selection, is also expected to improve the accuracy of predictions with regards to magic angles.

Since our computational approach relies on calculations within the single-particle picture, the TB band structures alone cannot provide quantitative insights into correlated behaviors observed in experimental measurements. Combining materials data and information from virtual (computer) and physical experiments into common knowledge bases is expected to improve the usefulness of our approach: imagine a synthesis tool that enables rapid fabrication of layered assemblies on a controlled manner with integrated {\em in situ} characterization, acting either in the role of the researcher or augmenting the knowledge base for more reliable predictions. A common practical concern is related to the synthesizability of materials predicted using modeling and simulation. Our atlas (Figure~\ref{fig:atlas}) provides a quick guide for picking promising candidates that are also structurally similar to known materials, making it easier to evaluate feasibility and needed resources for experimental verification. Moreover, in a Bayesian framework the knowledge obtained from unfruitful experiments can also be incorporated in the materials knowledge cycle by updating predictions, i.e., the probabilities $p(\theta)$, accordingly.

Our streamlined approach and specific findings provide insights into the electronic structure of twisted multi-layer graphene-based superlattices and offers reference for experimental measurements. It is straightforward to extend our computational strategy to explore other scientifically and technologically interesting 2D layered materials, including layered assemblies of transition metal dichalcogenides. Overall, this work aims to build confidence in the combined use of physics-based and data-driven modeling for the systematic refinement of knowledge about 2D layered materials, with implications for the development of novel quantum devices.

\section*{Acknowledgements}
G.A.T. would like to acknowledge Pavlos Protopapas and Ricardo Hausmann for stimulating discussions. This work was supported in part by the U.S. Department of Energy (DOE) Office of Science (Basic Energy Science) under Award No. DE-SC0019300 and by the U.S. National Science Foundation’s Science and Technology Center for Integrated Quantum Materials under Grant No. DMR-1231319. S.B.T. received support from the DOE Computational Sciences Graduate Fellowship (DOE CSGF) under grant number DE-FG02-97ER25308. 

Electronic structure calculations were performed on the Extreme Science and Engineering Discovery Environment (XSEDE), which is supported by National Science Foundation grant number ACI-1548562. Pilot calculations were performed on the Cannon cluster, supported by the FAS Division of Science Research Computing Group at Harvard University. We relied on computational resources of the National Energy Research Scientific Computing Center, a DOE facility operated under Contract No. DE-AC02-05CH11231, for the provisioning of databases supporting the computational work.

\bibliography{References}

\begin{thebibliography}{52}%
\makeatletter
\providecommand \@ifxundefined [1]{%
 \@ifx{#1\undefined}
}%
\providecommand \@ifnum [1]{%
 \ifnum #1\expandafter \@firstoftwo
 \else \expandafter \@secondoftwo
 \fi
}%
\providecommand \@ifx [1]{%
 \ifx #1\expandafter \@firstoftwo
 \else \expandafter \@secondoftwo
 \fi
}%
\providecommand \natexlab [1]{#1}%
\providecommand \enquote  [1]{``#1''}%
\providecommand \bibnamefont  [1]{#1}%
\providecommand \bibfnamefont [1]{#1}%
\providecommand \citenamefont [1]{#1}%
\providecommand \href@noop [0]{\@secondoftwo}%
\providecommand \href [0]{\begingroup \@sanitize@url \@href}%
\providecommand \@href[1]{\@@startlink{#1}\@@href}%
\providecommand \@@href[1]{\endgroup#1\@@endlink}%
\providecommand \@sanitize@url [0]{\catcode `\\12\catcode `\$12\catcode
  `\&12\catcode `\#12\catcode `\^12\catcode `\_12\catcode `\%12\relax}%
\providecommand \@@startlink[1]{}%
\providecommand \@@endlink[0]{}%
\providecommand \url  [0]{\begingroup\@sanitize@url \@url }%
\providecommand \@url [1]{\endgroup\@href {#1}{\urlprefix }}%
\providecommand \urlprefix  [0]{URL }%
\providecommand \Eprint [0]{\href }%
\providecommand \doibase [0]{http://dx.doi.org/}%
\providecommand \selectlanguage [0]{\@gobble}%
\providecommand \bibinfo  [0]{\@secondoftwo}%
\providecommand \bibfield  [0]{\@secondoftwo}%
\providecommand \translation [1]{[#1]}%
\providecommand \BibitemOpen [0]{}%
\providecommand \bibitemStop [0]{}%
\providecommand \bibitemNoStop [0]{.\EOS\space}%
\providecommand \EOS [0]{\spacefactor3000\relax}%
\providecommand \BibitemShut  [1]{\csname bibitem#1\endcsname}%
\let\auto@bib@innerbib\@empty
\bibitem [{\citenamefont {Geim}\ and\ \citenamefont
  {Grigorieva}(2013)}]{Geim2013Van}%
  \BibitemOpen
  \bibfield  {author} {\bibinfo {author} {\bibfnamefont {A.~K.}\ \bibnamefont
  {Geim}}\ and\ \bibinfo {author} {\bibfnamefont {I.~V.}\ \bibnamefont
  {Grigorieva}},\ }\href {\doibase 10.1038/nature12385} {\bibfield  {journal}
  {\bibinfo  {journal} {Nature}\ }\textbf {\bibinfo {volume} {499}},\ \bibinfo
  {pages} {419} (\bibinfo {year} {2013})}\BibitemShut {NoStop}%
\bibitem [{\citenamefont {Novoselov}\ \emph {et~al.}(2016)\citenamefont
  {Novoselov}, \citenamefont {Mishchenko}, \citenamefont {Carvalho},\ and\
  \citenamefont {Neto}}]{novoselov_2d_2016}%
  \BibitemOpen
  \bibfield  {author} {\bibinfo {author} {\bibfnamefont {K.~S.}\ \bibnamefont
  {Novoselov}}, \bibinfo {author} {\bibfnamefont {A.}~\bibnamefont
  {Mishchenko}}, \bibinfo {author} {\bibfnamefont {A.}~\bibnamefont
  {Carvalho}}, \ and\ \bibinfo {author} {\bibfnamefont {A.~H.~C.}\ \bibnamefont
  {Neto}},\ }\href {\doibase 10.1126/science.aac9439} {\bibfield  {journal}
  {\bibinfo  {journal} {Science}\ }\textbf {\bibinfo {volume} {353}} (\bibinfo
  {year} {2016}),\ 10.1126/science.aac9439}\BibitemShut {NoStop}%
\bibitem [{\citenamefont {Zhou}\ \emph {et~al.}(2018)\citenamefont {Zhou},
  \citenamefont {Lin}, \citenamefont {Huang}, \citenamefont {Zhou},
  \citenamefont {Chen}, \citenamefont {Xia}, \citenamefont {Wang},
  \citenamefont {Xie}, \citenamefont {Yu}, \citenamefont {Lei}, \citenamefont
  {Wu}, \citenamefont {Liu}, \citenamefont {Fu}, \citenamefont {Zeng},
  \citenamefont {Hsu}, \citenamefont {Yang}, \citenamefont {Lu}, \citenamefont
  {Yu}, \citenamefont {Shen}, \citenamefont {Lin}, \citenamefont {Yakobson},
  \citenamefont {Liu}, \citenamefont {Suenaga}, \citenamefont {Liu},\ and\
  \citenamefont {Liu}}]{Zhou2018library}%
  \BibitemOpen
  \bibfield  {author} {\bibinfo {author} {\bibfnamefont {J.}~\bibnamefont
  {Zhou}}, \bibinfo {author} {\bibfnamefont {J.}~\bibnamefont {Lin}}, \bibinfo
  {author} {\bibfnamefont {X.}~\bibnamefont {Huang}}, \bibinfo {author}
  {\bibfnamefont {Y.}~\bibnamefont {Zhou}}, \bibinfo {author} {\bibfnamefont
  {Y.}~\bibnamefont {Chen}}, \bibinfo {author} {\bibfnamefont {J.}~\bibnamefont
  {Xia}}, \bibinfo {author} {\bibfnamefont {H.}~\bibnamefont {Wang}}, \bibinfo
  {author} {\bibfnamefont {Y.}~\bibnamefont {Xie}}, \bibinfo {author}
  {\bibfnamefont {H.}~\bibnamefont {Yu}}, \bibinfo {author} {\bibfnamefont
  {J.}~\bibnamefont {Lei}}, \bibinfo {author} {\bibfnamefont {D.}~\bibnamefont
  {Wu}}, \bibinfo {author} {\bibfnamefont {F.}~\bibnamefont {Liu}}, \bibinfo
  {author} {\bibfnamefont {Q.}~\bibnamefont {Fu}}, \bibinfo {author}
  {\bibfnamefont {Q.}~\bibnamefont {Zeng}}, \bibinfo {author} {\bibfnamefont
  {C.-H.}\ \bibnamefont {Hsu}}, \bibinfo {author} {\bibfnamefont
  {C.}~\bibnamefont {Yang}}, \bibinfo {author} {\bibfnamefont {L.}~\bibnamefont
  {Lu}}, \bibinfo {author} {\bibfnamefont {T.}~\bibnamefont {Yu}}, \bibinfo
  {author} {\bibfnamefont {Z.}~\bibnamefont {Shen}}, \bibinfo {author}
  {\bibfnamefont {H.}~\bibnamefont {Lin}}, \bibinfo {author} {\bibfnamefont
  {B.~I.}\ \bibnamefont {Yakobson}}, \bibinfo {author} {\bibfnamefont
  {Q.}~\bibnamefont {Liu}}, \bibinfo {author} {\bibfnamefont {K.}~\bibnamefont
  {Suenaga}}, \bibinfo {author} {\bibfnamefont {G.}~\bibnamefont {Liu}}, \ and\
  \bibinfo {author} {\bibfnamefont {Z.}~\bibnamefont {Liu}},\ }\href {\doibase
  10.1038/s41586-018-0008-3} {\bibfield  {journal} {\bibinfo  {journal}
  {Nature}\ }\textbf {\bibinfo {volume} {556}},\ \bibinfo {pages} {355}
  (\bibinfo {year} {2018})}\BibitemShut {NoStop}%
\bibitem [{\citenamefont {Cheon}\ \emph {et~al.}(2017)\citenamefont {Cheon},
  \citenamefont {Duerloo}, \citenamefont {Sendek}, \citenamefont {Porter},
  \citenamefont {Chen},\ and\ \citenamefont {Reed}}]{Cheon2017Data}%
  \BibitemOpen
  \bibfield  {author} {\bibinfo {author} {\bibfnamefont {G.}~\bibnamefont
  {Cheon}}, \bibinfo {author} {\bibfnamefont {K.-A.~N.}\ \bibnamefont
  {Duerloo}}, \bibinfo {author} {\bibfnamefont {A.~D.}\ \bibnamefont {Sendek}},
  \bibinfo {author} {\bibfnamefont {C.}~\bibnamefont {Porter}}, \bibinfo
  {author} {\bibfnamefont {Y.}~\bibnamefont {Chen}}, \ and\ \bibinfo {author}
  {\bibfnamefont {E.~J.}\ \bibnamefont {Reed}},\ }\href {\doibase
  10.1021/acs.nanolett.6b05229} {\bibfield  {journal} {\bibinfo  {journal}
  {Nano Letters}\ }\textbf {\bibinfo {volume} {17}},\ \bibinfo {pages} {1915}
  (\bibinfo {year} {2017})}\BibitemShut {NoStop}%
\bibitem [{\citenamefont {Mounet}\ \emph {et~al.}(2018)\citenamefont {Mounet},
  \citenamefont {Gibertini}, \citenamefont {Schwaller}, \citenamefont {Campi},
  \citenamefont {Merkys}, \citenamefont {Marrazzo}, \citenamefont {Sohier},
  \citenamefont {Castelli}, \citenamefont {Cepellotti}, \citenamefont {Pizzi},\
  and\ \citenamefont {Marzari}}]{Mounet2018Two-dimensional}%
  \BibitemOpen
  \bibfield  {author} {\bibinfo {author} {\bibfnamefont {N.}~\bibnamefont
  {Mounet}}, \bibinfo {author} {\bibfnamefont {M.}~\bibnamefont {Gibertini}},
  \bibinfo {author} {\bibfnamefont {P.}~\bibnamefont {Schwaller}}, \bibinfo
  {author} {\bibfnamefont {D.}~\bibnamefont {Campi}}, \bibinfo {author}
  {\bibfnamefont {A.}~\bibnamefont {Merkys}}, \bibinfo {author} {\bibfnamefont
  {A.}~\bibnamefont {Marrazzo}}, \bibinfo {author} {\bibfnamefont
  {T.}~\bibnamefont {Sohier}}, \bibinfo {author} {\bibfnamefont {I.~E.}\
  \bibnamefont {Castelli}}, \bibinfo {author} {\bibfnamefont {A.}~\bibnamefont
  {Cepellotti}}, \bibinfo {author} {\bibfnamefont {G.}~\bibnamefont {Pizzi}}, \
  and\ \bibinfo {author} {\bibfnamefont {N.}~\bibnamefont {Marzari}},\ }\href
  {\doibase 10.1038/s41565-017-0035-5} {\bibfield  {journal} {\bibinfo
  {journal} {Nature Nanotechnology}\ }\textbf {\bibinfo {volume} {13}},\
  \bibinfo {pages} {246} (\bibinfo {year} {2018})}\BibitemShut {NoStop}%
\bibitem [{\citenamefont {Carr}\ \emph {et~al.}(2017)\citenamefont {Carr},
  \citenamefont {Massatt}, \citenamefont {Fang}, \citenamefont {Cazeaux},
  \citenamefont {Luskin},\ and\ \citenamefont
  {Kaxiras}}]{Carr2017Twistronics:}%
  \BibitemOpen
  \bibfield  {author} {\bibinfo {author} {\bibfnamefont {S.}~\bibnamefont
  {Carr}}, \bibinfo {author} {\bibfnamefont {D.}~\bibnamefont {Massatt}},
  \bibinfo {author} {\bibfnamefont {S.}~\bibnamefont {Fang}}, \bibinfo {author}
  {\bibfnamefont {P.}~\bibnamefont {Cazeaux}}, \bibinfo {author} {\bibfnamefont
  {M.}~\bibnamefont {Luskin}}, \ and\ \bibinfo {author} {\bibfnamefont
  {E.}~\bibnamefont {Kaxiras}},\ }\href {\doibase 10.1103/PhysRevB.95.075420}
  {\bibfield  {journal} {\bibinfo  {journal} {Physical Review B}\ }\textbf
  {\bibinfo {volume} {95}},\ \bibinfo {pages} {075420} (\bibinfo {year}
  {2017})}\BibitemShut {NoStop}%
\bibitem [{\citenamefont {Cao}\ \emph {et~al.}(2018{\natexlab{a}})\citenamefont
  {Cao}, \citenamefont {Fatemi}, \citenamefont {Fang}, \citenamefont
  {Watanabe}, \citenamefont {Taniguchi}, \citenamefont {Kaxiras},\ and\
  \citenamefont {Jarillo-Herrero}}]{Cao2018Unconventional}%
  \BibitemOpen
  \bibfield  {author} {\bibinfo {author} {\bibfnamefont {Y.}~\bibnamefont
  {Cao}}, \bibinfo {author} {\bibfnamefont {V.}~\bibnamefont {Fatemi}},
  \bibinfo {author} {\bibfnamefont {S.}~\bibnamefont {Fang}}, \bibinfo {author}
  {\bibfnamefont {K.}~\bibnamefont {Watanabe}}, \bibinfo {author}
  {\bibfnamefont {T.}~\bibnamefont {Taniguchi}}, \bibinfo {author}
  {\bibfnamefont {E.}~\bibnamefont {Kaxiras}}, \ and\ \bibinfo {author}
  {\bibfnamefont {P.}~\bibnamefont {Jarillo-Herrero}},\ }\href {\doibase
  10.1038/nature26160} {\bibfield  {journal} {\bibinfo  {journal} {Nature}\
  }\textbf {\bibinfo {volume} {556}},\ \bibinfo {pages} {43} (\bibinfo {year}
  {2018}{\natexlab{a}})}\BibitemShut {NoStop}%
\bibitem [{\citenamefont {Cao}\ \emph {et~al.}(2018{\natexlab{b}})\citenamefont
  {Cao}, \citenamefont {Fatemi}, \citenamefont {Demir}, \citenamefont {Fang},
  \citenamefont {Tomarken}, \citenamefont {Luo}, \citenamefont
  {Sanchez-Yamagishi}, \citenamefont {Watanabe}, \citenamefont {Taniguchi},
  \citenamefont {Kaxiras}, \citenamefont {Ashoori},\ and\ \citenamefont
  {Jarillo-Herrero}}]{Cao2018Correlated}%
  \BibitemOpen
  \bibfield  {author} {\bibinfo {author} {\bibfnamefont {Y.}~\bibnamefont
  {Cao}}, \bibinfo {author} {\bibfnamefont {V.}~\bibnamefont {Fatemi}},
  \bibinfo {author} {\bibfnamefont {A.}~\bibnamefont {Demir}}, \bibinfo
  {author} {\bibfnamefont {S.}~\bibnamefont {Fang}}, \bibinfo {author}
  {\bibfnamefont {S.~L.}\ \bibnamefont {Tomarken}}, \bibinfo {author}
  {\bibfnamefont {J.~Y.}\ \bibnamefont {Luo}}, \bibinfo {author} {\bibfnamefont
  {J.~D.}\ \bibnamefont {Sanchez-Yamagishi}}, \bibinfo {author} {\bibfnamefont
  {K.}~\bibnamefont {Watanabe}}, \bibinfo {author} {\bibfnamefont
  {T.}~\bibnamefont {Taniguchi}}, \bibinfo {author} {\bibfnamefont
  {E.}~\bibnamefont {Kaxiras}}, \bibinfo {author} {\bibfnamefont {R.~C.}\
  \bibnamefont {Ashoori}}, \ and\ \bibinfo {author} {\bibfnamefont
  {P.}~\bibnamefont {Jarillo-Herrero}},\ }\href {\doibase 10.1038/nature26154}
  {\bibfield  {journal} {\bibinfo  {journal} {Nature}\ }\textbf {\bibinfo
  {volume} {556}},\ \bibinfo {pages} {80} (\bibinfo {year}
  {2018}{\natexlab{b}})}\BibitemShut {NoStop}%
\bibitem [{\citenamefont {Bistritzer}\ and\ \citenamefont
  {MacDonald}(2011)}]{Bistritzer2011Moire}%
  \BibitemOpen
  \bibfield  {author} {\bibinfo {author} {\bibfnamefont {R.}~\bibnamefont
  {Bistritzer}}\ and\ \bibinfo {author} {\bibfnamefont {A.~H.}\ \bibnamefont
  {MacDonald}},\ }\href {\doibase 10.1073/pnas.1108174108} {\bibfield
  {journal} {\bibinfo  {journal} {Proceedings of the National Academy of
  Sciences}\ }\textbf {\bibinfo {volume} {108}},\ \bibinfo {pages} {12233}
  (\bibinfo {year} {2011})}\BibitemShut {NoStop}%
\bibitem [{\citenamefont {Liu}\ \emph {et~al.}(2019)\citenamefont {Liu},
  \citenamefont {Hao}, \citenamefont {Khalaf}, \citenamefont {Lee},
  \citenamefont {Watanabe}, \citenamefont {Taniguchi}, \citenamefont
  {Vishwanath},\ and\ \citenamefont {Kim}}]{Liu2019Spin-polarized}%
  \BibitemOpen
  \bibfield  {author} {\bibinfo {author} {\bibfnamefont {X.}~\bibnamefont
  {Liu}}, \bibinfo {author} {\bibfnamefont {Z.}~\bibnamefont {Hao}}, \bibinfo
  {author} {\bibfnamefont {E.}~\bibnamefont {Khalaf}}, \bibinfo {author}
  {\bibfnamefont {J.~Y.}\ \bibnamefont {Lee}}, \bibinfo {author} {\bibfnamefont
  {K.}~\bibnamefont {Watanabe}}, \bibinfo {author} {\bibfnamefont
  {T.}~\bibnamefont {Taniguchi}}, \bibinfo {author} {\bibfnamefont
  {A.}~\bibnamefont {Vishwanath}}, \ and\ \bibinfo {author} {\bibfnamefont
  {P.}~\bibnamefont {Kim}},\ }\href {http://arxiv.org/abs/1903.08130}
  {\bibfield  {journal} {\bibinfo  {journal} {arXiv:1903.08130 [cond-mat]}\ }
  (\bibinfo {year} {2019})},\ \bibinfo {note} {arXiv: 1903.08130}\BibitemShut
  {NoStop}%
\bibitem [{\citenamefont {Su{\'a}rez~Morell}\ \emph {et~al.}(2013)\citenamefont
  {Su{\'a}rez~Morell}, \citenamefont {Pacheco}, \citenamefont {Chico},\ and\
  \citenamefont {Brey}}]{SuarezMorell2013Electronic}%
  \BibitemOpen
  \bibfield  {author} {\bibinfo {author} {\bibfnamefont {E.}~\bibnamefont
  {Su{\'a}rez~Morell}}, \bibinfo {author} {\bibfnamefont {M.}~\bibnamefont
  {Pacheco}}, \bibinfo {author} {\bibfnamefont {L.}~\bibnamefont {Chico}}, \
  and\ \bibinfo {author} {\bibfnamefont {L.}~\bibnamefont {Brey}},\ }\href
  {\doibase 10.1103/PhysRevB.87.125414} {\bibfield  {journal} {\bibinfo
  {journal} {Physical Review B}\ }\textbf {\bibinfo {volume} {87}},\ \bibinfo
  {pages} {125414} (\bibinfo {year} {2013})}\BibitemShut {NoStop}%
\bibitem [{\citenamefont {Tritsaris}\ and\ \citenamefont
  {Rossmeisl}(2012)}]{Tritsaris2012Methanol}%
  \BibitemOpen
  \bibfield  {author} {\bibinfo {author} {\bibfnamefont {G.~A.}\ \bibnamefont
  {Tritsaris}}\ and\ \bibinfo {author} {\bibfnamefont {J.}~\bibnamefont
  {Rossmeisl}},\ }\href {\doibase 10.1021/jp209506d} {\bibfield  {journal}
  {\bibinfo  {journal} {The Journal of Physical Chemistry C}\ }\textbf
  {\bibinfo {volume} {116}},\ \bibinfo {pages} {11980} (\bibinfo {year}
  {2012})}\BibitemShut {NoStop}%
\bibitem [{\citenamefont {Chen}\ \emph {et~al.}(2016)\citenamefont {Chen},
  \citenamefont {P\"{o}hls}, \citenamefont {Hautier}, \citenamefont {Broberg},
  \citenamefont {Bajaj}, \citenamefont {Aydemir}, \citenamefont {Gibbs},
  \citenamefont {Zhu}, \citenamefont {Asta}, \citenamefont {Snyder},
  \citenamefont {Meredig}, \citenamefont {White}, \citenamefont {Persson},\
  and\ \citenamefont {Jain}}]{Chen2016Understanding}%
  \BibitemOpen
  \bibfield  {author} {\bibinfo {author} {\bibfnamefont {W.}~\bibnamefont
  {Chen}}, \bibinfo {author} {\bibfnamefont {J.-H.}\ \bibnamefont {P\"{o}hls}},
  \bibinfo {author} {\bibfnamefont {G.}~\bibnamefont {Hautier}}, \bibinfo
  {author} {\bibfnamefont {D.}~\bibnamefont {Broberg}}, \bibinfo {author}
  {\bibfnamefont {S.}~\bibnamefont {Bajaj}}, \bibinfo {author} {\bibfnamefont
  {U.}~\bibnamefont {Aydemir}}, \bibinfo {author} {\bibfnamefont {Z.~M.}\
  \bibnamefont {Gibbs}}, \bibinfo {author} {\bibfnamefont {H.}~\bibnamefont
  {Zhu}}, \bibinfo {author} {\bibfnamefont {M.}~\bibnamefont {Asta}}, \bibinfo
  {author} {\bibfnamefont {G.~J.}\ \bibnamefont {Snyder}}, \bibinfo {author}
  {\bibfnamefont {B.}~\bibnamefont {Meredig}}, \bibinfo {author} {\bibfnamefont
  {M.~A.}\ \bibnamefont {White}}, \bibinfo {author} {\bibfnamefont
  {K.}~\bibnamefont {Persson}}, \ and\ \bibinfo {author} {\bibfnamefont
  {A.}~\bibnamefont {Jain}},\ }\href {\doibase 10.1039/C5TC04339E} {\bibfield
  {journal} {\bibinfo  {journal} {Journal of Materials Chemistry C}\ }\textbf
  {\bibinfo {volume} {4}},\ \bibinfo {pages} {4414} (\bibinfo {year}
  {2016})}\BibitemShut {NoStop}%
\bibitem [{\citenamefont {J{\'o}hannesson}\ \emph {et~al.}(2002)\citenamefont
  {J{\'o}hannesson}, \citenamefont {Bligaard}, \citenamefont {Ruban},
  \citenamefont {Skriver}, \citenamefont {Jacobsen},\ and\ \citenamefont
  {N{\o}rskov}}]{Johannesson2002Combined}%
  \BibitemOpen
  \bibfield  {author} {\bibinfo {author} {\bibfnamefont {G.}~\bibnamefont
  {J{\'o}hannesson}}, \bibinfo {author} {\bibfnamefont {T.}~\bibnamefont
  {Bligaard}}, \bibinfo {author} {\bibfnamefont {A.}~\bibnamefont {Ruban}},
  \bibinfo {author} {\bibfnamefont {H.}~\bibnamefont {Skriver}}, \bibinfo
  {author} {\bibfnamefont {K.}~\bibnamefont {Jacobsen}}, \ and\ \bibinfo
  {author} {\bibfnamefont {J.}~\bibnamefont {N{\o}rskov}},\ }\href@noop {}
  {\bibfield  {journal} {\bibinfo  {journal} {Physical Review Letters}\
  }\textbf {\bibinfo {volume} {88}},\ \bibinfo {pages} {2555061} (\bibinfo
  {year} {2002})}\BibitemShut {NoStop}%
\bibitem [{\citenamefont {Torrisi}\ \emph {et~al.}(2019)\citenamefont
  {Torrisi}, \citenamefont {Singh}, \citenamefont {Montoya},\ and\
  \citenamefont {Persson}}]{Torrisi2019}%
  \BibitemOpen
  \bibfield  {author} {\bibinfo {author} {\bibfnamefont {S.}~\bibnamefont
  {Torrisi}}, \bibinfo {author} {\bibfnamefont {A.}~\bibnamefont {Singh}},
  \bibinfo {author} {\bibfnamefont {J.}~\bibnamefont {Montoya}}, \ and\
  \bibinfo {author} {\bibfnamefont {K.}~\bibnamefont {Persson}},\ }\href@noop
  {} {\bibfield  {journal} {\bibinfo  {journal} {arXiv}\ } (\bibinfo {year}
  {2019})},\ \Eprint {http://arxiv.org/abs/1912.09545v1} {arXiv:1912.09545v1}
  \BibitemShut {NoStop}%
\bibitem [{\citenamefont {Tran}\ and\ \citenamefont
  {Ulissi}(2018)}]{tran_active_2018}%
  \BibitemOpen
  \bibfield  {author} {\bibinfo {author} {\bibfnamefont {K.}~\bibnamefont
  {Tran}}\ and\ \bibinfo {author} {\bibfnamefont {Z.~W.}\ \bibnamefont
  {Ulissi}},\ }\href {\doibase 10.1038/s41929-018-0142-1} {\bibfield  {journal}
  {\bibinfo  {journal} {Nature Catalysis}\ }\textbf {\bibinfo {volume} {1}},\
  \bibinfo {pages} {696} (\bibinfo {year} {2018})}\BibitemShut {NoStop}%
\bibitem [{\citenamefont {Bassman}\ \emph {et~al.}(2018)\citenamefont
  {Bassman}, \citenamefont {Rajak}, \citenamefont {Kalia}, \citenamefont
  {Nakano}, \citenamefont {Sha}, \citenamefont {Sun}, \citenamefont {Singh},
  \citenamefont {Aykol}, \citenamefont {Huck}, \citenamefont {Persson},\ and\
  \citenamefont {Vashishta}}]{Bassman2018Active}%
  \BibitemOpen
  \bibfield  {author} {\bibinfo {author} {\bibfnamefont {L.}~\bibnamefont
  {Bassman}}, \bibinfo {author} {\bibfnamefont {P.}~\bibnamefont {Rajak}},
  \bibinfo {author} {\bibfnamefont {R.~K.}\ \bibnamefont {Kalia}}, \bibinfo
  {author} {\bibfnamefont {A.}~\bibnamefont {Nakano}}, \bibinfo {author}
  {\bibfnamefont {F.}~\bibnamefont {Sha}}, \bibinfo {author} {\bibfnamefont
  {J.}~\bibnamefont {Sun}}, \bibinfo {author} {\bibfnamefont {D.~J.}\
  \bibnamefont {Singh}}, \bibinfo {author} {\bibfnamefont {M.}~\bibnamefont
  {Aykol}}, \bibinfo {author} {\bibfnamefont {P.}~\bibnamefont {Huck}},
  \bibinfo {author} {\bibfnamefont {K.}~\bibnamefont {Persson}}, \ and\
  \bibinfo {author} {\bibfnamefont {P.}~\bibnamefont {Vashishta}},\ }\href
  {\doibase 10.1038/s41524-018-0129-0} {\bibfield  {journal} {\bibinfo
  {journal} {npj Computational Materials}\ }\textbf {\bibinfo {volume} {4}},\
  \bibinfo {pages} {74} (\bibinfo {year} {2018})}\BibitemShut {NoStop}%
\bibitem [{\citenamefont {Haastrup}\ \emph {et~al.}(2018)\citenamefont
  {Haastrup}, \citenamefont {Strange}, \citenamefont {Pandey}, \citenamefont
  {Deilmann}, \citenamefont {Schmidt}, \citenamefont {Hinsche}, \citenamefont
  {Gjerding}, \citenamefont {Torelli}, \citenamefont {Larsen}, \citenamefont
  {Riis-Jensen}, \citenamefont {Gath}, \citenamefont {Jacobsen}, \citenamefont
  {Mortensen}, \citenamefont {Olsen},\ and\ \citenamefont
  {Thygesen}}]{Haastrup2018Computational}%
  \BibitemOpen
  \bibfield  {author} {\bibinfo {author} {\bibfnamefont {S.}~\bibnamefont
  {Haastrup}}, \bibinfo {author} {\bibfnamefont {M.}~\bibnamefont {Strange}},
  \bibinfo {author} {\bibfnamefont {M.}~\bibnamefont {Pandey}}, \bibinfo
  {author} {\bibfnamefont {T.}~\bibnamefont {Deilmann}}, \bibinfo {author}
  {\bibfnamefont {P.~S.}\ \bibnamefont {Schmidt}}, \bibinfo {author}
  {\bibfnamefont {N.~F.}\ \bibnamefont {Hinsche}}, \bibinfo {author}
  {\bibfnamefont {M.~N.}\ \bibnamefont {Gjerding}}, \bibinfo {author}
  {\bibfnamefont {D.}~\bibnamefont {Torelli}}, \bibinfo {author} {\bibfnamefont
  {P.~M.}\ \bibnamefont {Larsen}}, \bibinfo {author} {\bibfnamefont {A.~C.}\
  \bibnamefont {Riis-Jensen}}, \bibinfo {author} {\bibfnamefont
  {J.}~\bibnamefont {Gath}}, \bibinfo {author} {\bibfnamefont {K.~W.}\
  \bibnamefont {Jacobsen}}, \bibinfo {author} {\bibfnamefont {J.~J.}\
  \bibnamefont {Mortensen}}, \bibinfo {author} {\bibfnamefont {T.}~\bibnamefont
  {Olsen}}, \ and\ \bibinfo {author} {\bibfnamefont {K.~S.}\ \bibnamefont
  {Thygesen}},\ }\href {\doibase 10.1088/2053-1583/aacfc1} {\bibfield
  {journal} {\bibinfo  {journal} {2D Materials}\ }\textbf {\bibinfo {volume}
  {5}},\ \bibinfo {pages} {042002} (\bibinfo {year} {2018})}\BibitemShut
  {NoStop}%
\bibitem [{\citenamefont {Tritsaris}\ \emph {et~al.}(2016)\citenamefont
  {Tritsaris}, \citenamefont {Shirodkar}, \citenamefont {Kaxiras},
  \citenamefont {Cazeaux}, \citenamefont {Luskin}, \citenamefont {Plecháč},\
  and\ \citenamefont {Cancès}}]{Tritsaris2016Perturbation}%
  \BibitemOpen
  \bibfield  {author} {\bibinfo {author} {\bibfnamefont {G.~A.}\ \bibnamefont
  {Tritsaris}}, \bibinfo {author} {\bibfnamefont {S.~N.}\ \bibnamefont
  {Shirodkar}}, \bibinfo {author} {\bibfnamefont {E.}~\bibnamefont {Kaxiras}},
  \bibinfo {author} {\bibfnamefont {P.}~\bibnamefont {Cazeaux}}, \bibinfo
  {author} {\bibfnamefont {M.}~\bibnamefont {Luskin}}, \bibinfo {author}
  {\bibfnamefont {P.}~\bibnamefont {Plecháč}}, \ and\ \bibinfo {author}
  {\bibfnamefont {E.}~\bibnamefont {Cancès}},\ }\href {\doibase
  10.1557/jmr.2016.99} {\bibfield  {journal} {\bibinfo  {journal} {Journal of
  Materials Research}\ }\textbf {\bibinfo {volume} {31}},\ \bibinfo {pages}
  {959} (\bibinfo {year} {2016})}\BibitemShut {NoStop}%
\bibitem [{\citenamefont {Fang}\ and\ \citenamefont
  {Kaxiras}(2016)}]{Fang2016Electronic}%
  \BibitemOpen
  \bibfield  {author} {\bibinfo {author} {\bibfnamefont {S.}~\bibnamefont
  {Fang}}\ and\ \bibinfo {author} {\bibfnamefont {E.}~\bibnamefont {Kaxiras}},\
  }\href {\doibase 10.1103/PhysRevB.93.235153} {\bibfield  {journal} {\bibinfo
  {journal} {Physical Review B}\ }\textbf {\bibinfo {volume} {93}},\ \bibinfo
  {pages} {235153} (\bibinfo {year} {2016})}\BibitemShut {NoStop}%
\bibitem [{\citenamefont {Jain}\ \emph {et~al.}(2015)\citenamefont {Jain},
  \citenamefont {Ong}, \citenamefont {Chen}, \citenamefont {Medasani},
  \citenamefont {Qu}, \citenamefont {Kocher}, \citenamefont {Brafman},
  \citenamefont {Petretto}, \citenamefont {Rignanese}, \citenamefont {Hautier},
  \citenamefont {Gunter},\ and\ \citenamefont {Persson}}]{Jain2015FireWorks:}%
  \BibitemOpen
  \bibfield  {author} {\bibinfo {author} {\bibfnamefont {A.}~\bibnamefont
  {Jain}}, \bibinfo {author} {\bibfnamefont {S.~P.}\ \bibnamefont {Ong}},
  \bibinfo {author} {\bibfnamefont {W.}~\bibnamefont {Chen}}, \bibinfo {author}
  {\bibfnamefont {B.}~\bibnamefont {Medasani}}, \bibinfo {author}
  {\bibfnamefont {X.}~\bibnamefont {Qu}}, \bibinfo {author} {\bibfnamefont
  {M.}~\bibnamefont {Kocher}}, \bibinfo {author} {\bibfnamefont
  {M.}~\bibnamefont {Brafman}}, \bibinfo {author} {\bibfnamefont
  {G.}~\bibnamefont {Petretto}}, \bibinfo {author} {\bibfnamefont {G.-M.}\
  \bibnamefont {Rignanese}}, \bibinfo {author} {\bibfnamefont {G.}~\bibnamefont
  {Hautier}}, \bibinfo {author} {\bibfnamefont {D.}~\bibnamefont {Gunter}}, \
  and\ \bibinfo {author} {\bibfnamefont {K.~A.}\ \bibnamefont {Persson}},\
  }\href {\doibase 10.1002/cpe.3505} {\bibfield  {journal} {\bibinfo  {journal}
  {Concurrency and Computation: Practice and Experience}\ }\textbf {\bibinfo
  {volume} {27}},\ \bibinfo {pages} {5037} (\bibinfo {year}
  {2015})}\BibitemShut {NoStop}%
\bibitem [{\citenamefont {Curtarolo}\ \emph {et~al.}(2012)\citenamefont
  {Curtarolo}, \citenamefont {Setyawan}, \citenamefont {Hart}, \citenamefont
  {Jahnatek}, \citenamefont {Chepulskii}, \citenamefont {Taylor}, \citenamefont
  {Wang}, \citenamefont {Xue}, \citenamefont {Yang}, \citenamefont {Levy},
  \citenamefont {Mehl}, \citenamefont {Stokes}, \citenamefont {Demchenko},\
  and\ \citenamefont {Morgan}}]{Curtarolo2012AFLOW:}%
  \BibitemOpen
  \bibfield  {author} {\bibinfo {author} {\bibfnamefont {S.}~\bibnamefont
  {Curtarolo}}, \bibinfo {author} {\bibfnamefont {W.}~\bibnamefont {Setyawan}},
  \bibinfo {author} {\bibfnamefont {G.~L.~W.}\ \bibnamefont {Hart}}, \bibinfo
  {author} {\bibfnamefont {M.}~\bibnamefont {Jahnatek}}, \bibinfo {author}
  {\bibfnamefont {R.~V.}\ \bibnamefont {Chepulskii}}, \bibinfo {author}
  {\bibfnamefont {R.~H.}\ \bibnamefont {Taylor}}, \bibinfo {author}
  {\bibfnamefont {S.}~\bibnamefont {Wang}}, \bibinfo {author} {\bibfnamefont
  {J.}~\bibnamefont {Xue}}, \bibinfo {author} {\bibfnamefont {K.}~\bibnamefont
  {Yang}}, \bibinfo {author} {\bibfnamefont {O.}~\bibnamefont {Levy}}, \bibinfo
  {author} {\bibfnamefont {M.~J.}\ \bibnamefont {Mehl}}, \bibinfo {author}
  {\bibfnamefont {H.~T.}\ \bibnamefont {Stokes}}, \bibinfo {author}
  {\bibfnamefont {D.~O.}\ \bibnamefont {Demchenko}}, \ and\ \bibinfo {author}
  {\bibfnamefont {D.}~\bibnamefont {Morgan}},\ }\href {\doibase
  10.1016/j.commatsci.2012.02.005} {\bibfield  {journal} {\bibinfo  {journal}
  {Computational Materials Science}\ }\textbf {\bibinfo {volume} {58}},\
  \bibinfo {pages} {218} (\bibinfo {year} {2012})}\BibitemShut {NoStop}%
\bibitem [{\citenamefont {Pizzi}\ \emph {et~al.}(2016)\citenamefont {Pizzi},
  \citenamefont {Cepellotti}, \citenamefont {Sabatini}, \citenamefont
  {Marzari},\ and\ \citenamefont {Kozinsky}}]{Pizzi2016AiiDA:}%
  \BibitemOpen
  \bibfield  {author} {\bibinfo {author} {\bibfnamefont {G.}~\bibnamefont
  {Pizzi}}, \bibinfo {author} {\bibfnamefont {A.}~\bibnamefont {Cepellotti}},
  \bibinfo {author} {\bibfnamefont {R.}~\bibnamefont {Sabatini}}, \bibinfo
  {author} {\bibfnamefont {N.}~\bibnamefont {Marzari}}, \ and\ \bibinfo
  {author} {\bibfnamefont {B.}~\bibnamefont {Kozinsky}},\ }\href {\doibase
  10.1016/j.commatsci.2015.09.013} {\bibfield  {journal} {\bibinfo  {journal}
  {Computational Materials Science}\ }\textbf {\bibinfo {volume} {111}},\
  \bibinfo {pages} {218} (\bibinfo {year} {2016})}\BibitemShut {NoStop}%
\bibitem [{\citenamefont {Sun}\ \emph {et~al.}(2019)\citenamefont {Sun},
  \citenamefont {Hartono}, \citenamefont {Ren}, \citenamefont {Oviedo},
  \citenamefont {Buscemi}, \citenamefont {Layurova}, \citenamefont {Chen},
  \citenamefont {Ogunfunmi}, \citenamefont {Thapa}, \citenamefont {Ramasamy},
  \citenamefont {Settens}, \citenamefont {DeCost}, \citenamefont {Kusne},
  \citenamefont {Liu}, \citenamefont {Tian}, \citenamefont {Peters},
  \citenamefont {Correa-Baena},\ and\ \citenamefont
  {Buonassisi}}]{Sun2019Accelerated}%
  \BibitemOpen
  \bibfield  {author} {\bibinfo {author} {\bibfnamefont {S.}~\bibnamefont
  {Sun}}, \bibinfo {author} {\bibfnamefont {N.~T.~P.}\ \bibnamefont {Hartono}},
  \bibinfo {author} {\bibfnamefont {Z.~D.}\ \bibnamefont {Ren}}, \bibinfo
  {author} {\bibfnamefont {F.}~\bibnamefont {Oviedo}}, \bibinfo {author}
  {\bibfnamefont {A.~M.}\ \bibnamefont {Buscemi}}, \bibinfo {author}
  {\bibfnamefont {M.}~\bibnamefont {Layurova}}, \bibinfo {author}
  {\bibfnamefont {D.~X.}\ \bibnamefont {Chen}}, \bibinfo {author}
  {\bibfnamefont {T.}~\bibnamefont {Ogunfunmi}}, \bibinfo {author}
  {\bibfnamefont {J.}~\bibnamefont {Thapa}}, \bibinfo {author} {\bibfnamefont
  {S.}~\bibnamefont {Ramasamy}}, \bibinfo {author} {\bibfnamefont
  {C.}~\bibnamefont {Settens}}, \bibinfo {author} {\bibfnamefont {B.~L.}\
  \bibnamefont {DeCost}}, \bibinfo {author} {\bibfnamefont {A.~G.}\
  \bibnamefont {Kusne}}, \bibinfo {author} {\bibfnamefont {Z.}~\bibnamefont
  {Liu}}, \bibinfo {author} {\bibfnamefont {S.~I.~P.}\ \bibnamefont {Tian}},
  \bibinfo {author} {\bibfnamefont {I.~M.}\ \bibnamefont {Peters}}, \bibinfo
  {author} {\bibfnamefont {J.-P.}\ \bibnamefont {Correa-Baena}}, \ and\
  \bibinfo {author} {\bibfnamefont {T.}~\bibnamefont {Buonassisi}},\ }\href
  {\doibase 10.1016/j.joule.2019.05.014} {\bibfield  {journal} {\bibinfo
  {journal} {Joule}\ }\textbf {\bibinfo {volume} {3}},\ \bibinfo {pages} {1437}
  (\bibinfo {year} {2019})}\BibitemShut {NoStop}%
\bibitem [{\citenamefont {Botu}\ and\ \citenamefont
  {Ramprasad}(2015)}]{Botu2015Adaptive}%
  \BibitemOpen
  \bibfield  {author} {\bibinfo {author} {\bibfnamefont {V.}~\bibnamefont
  {Botu}}\ and\ \bibinfo {author} {\bibfnamefont {R.}~\bibnamefont
  {Ramprasad}},\ }\href {\doibase 10.1002/qua.24836} {\bibfield  {journal}
  {\bibinfo  {journal} {International Journal of Quantum Chemistry}\ }\textbf
  {\bibinfo {volume} {115}},\ \bibinfo {pages} {1074} (\bibinfo {year}
  {2015})}\BibitemShut {NoStop}%
\bibitem [{\citenamefont {Cubuk}\ \emph {et~al.}(2015)\citenamefont {Cubuk},
  \citenamefont {Schoenholz}, \citenamefont {Rieser}, \citenamefont {Malone},
  \citenamefont {Rottler}, \citenamefont {Durian}, \citenamefont {Kaxiras},\
  and\ \citenamefont {Liu}}]{Cubuk2015Identifying}%
  \BibitemOpen
  \bibfield  {author} {\bibinfo {author} {\bibfnamefont {E.}~\bibnamefont
  {Cubuk}}, \bibinfo {author} {\bibfnamefont {S.}~\bibnamefont {Schoenholz}},
  \bibinfo {author} {\bibfnamefont {J.}~\bibnamefont {Rieser}}, \bibinfo
  {author} {\bibfnamefont {B.}~\bibnamefont {Malone}}, \bibinfo {author}
  {\bibfnamefont {J.}~\bibnamefont {Rottler}}, \bibinfo {author} {\bibfnamefont
  {D.}~\bibnamefont {Durian}}, \bibinfo {author} {\bibfnamefont
  {E.}~\bibnamefont {Kaxiras}}, \ and\ \bibinfo {author} {\bibfnamefont
  {A.}~\bibnamefont {Liu}},\ }\href {\doibase 10.1103/PhysRevLett.114.108001}
  {\bibfield  {journal} {\bibinfo  {journal} {Physical Review Letters}\
  }\textbf {\bibinfo {volume} {114}} (\bibinfo {year} {2015}),\
  10.1103/PhysRevLett.114.108001},\ \bibinfo {note} {[Online; accessed
  2016-12-11]}\BibitemShut {NoStop}%
\bibitem [{\citenamefont {Janet}\ \emph {et~al.}(2019)\citenamefont {Janet},
  \citenamefont {Duan}, \citenamefont {Yang}, \citenamefont {Nandy},\ and\
  \citenamefont {Kulik}}]{Janet2019quantitative}%
  \BibitemOpen
  \bibfield  {author} {\bibinfo {author} {\bibfnamefont {J.~P.}\ \bibnamefont
  {Janet}}, \bibinfo {author} {\bibfnamefont {C.}~\bibnamefont {Duan}},
  \bibinfo {author} {\bibfnamefont {T.}~\bibnamefont {Yang}}, \bibinfo {author}
  {\bibfnamefont {A.}~\bibnamefont {Nandy}}, \ and\ \bibinfo {author}
  {\bibfnamefont {H.~J.}\ \bibnamefont {Kulik}},\ }\href {\doibase
  10.1039/C9SC02298H} {\bibfield  {journal} {\bibinfo  {journal} {Chemical
  Science}\ } (\bibinfo {year} {2019}),\ 10.1039/C9SC02298H},\ \bibinfo {note}
  {[Online; accessed 2019-08-12]}\BibitemShut {NoStop}%
\bibitem [{\citenamefont {Sparks}\ \emph {et~al.}(2016)\citenamefont {Sparks},
  \citenamefont {Gaultois}, \citenamefont {Oliynyk}, \citenamefont {Brgoch},\
  and\ \citenamefont {Meredig}}]{Sparks2016Data}%
  \BibitemOpen
  \bibfield  {author} {\bibinfo {author} {\bibfnamefont {T.~D.}\ \bibnamefont
  {Sparks}}, \bibinfo {author} {\bibfnamefont {M.~W.}\ \bibnamefont
  {Gaultois}}, \bibinfo {author} {\bibfnamefont {A.}~\bibnamefont {Oliynyk}},
  \bibinfo {author} {\bibfnamefont {J.}~\bibnamefont {Brgoch}}, \ and\ \bibinfo
  {author} {\bibfnamefont {B.}~\bibnamefont {Meredig}},\ }\href {\doibase
  10.1016/j.scriptamat.2015.04.026} {\bibfield  {journal} {\bibinfo  {journal}
  {Scripta Materialia}\ }\bibinfo {series} {Viewpoint Set No. 57: Contemporary
  Innovations for Thermoelectrics Research and Development},\ \textbf {\bibinfo
  {volume} {111}},\ \bibinfo {pages} {10} (\bibinfo {year} {2016})}\BibitemShut
  {NoStop}%
\bibitem [{\citenamefont {Liao}(2005)}]{Shu-HsienLiao2005Expert}%
  \BibitemOpen
  \bibfield  {author} {\bibinfo {author} {\bibfnamefont {S.-H.}\ \bibnamefont
  {Liao}},\ }\href {\doibase 10.1016/j.eswa.2004.08.003} {\bibfield  {journal}
  {\bibinfo  {journal} {Expert Systems with Applications}\ }\textbf {\bibinfo
  {volume} {28}},\ \bibinfo {pages} {93} (\bibinfo {year} {2005})}\BibitemShut
  {NoStop}%
\bibitem [{\citenamefont {Turban}\ \emph {et~al.}(2004)\citenamefont {Turban},
  \citenamefont {Aronson},\ and\ \citenamefont {Liang}}]{Turban2004Decision}%
  \BibitemOpen
  \bibfield  {author} {\bibinfo {author} {\bibfnamefont {E.}~\bibnamefont
  {Turban}}, \bibinfo {author} {\bibfnamefont {J.~E.}\ \bibnamefont {Aronson}},
  \ and\ \bibinfo {author} {\bibfnamefont {T.-P.}\ \bibnamefont {Liang}},\
  }\href@noop {} {\emph {\bibinfo {title} {Decision Support Systems and
  Intelligent Systems}}},\ \bibinfo {edition} {7th}\ ed.\ (\bibinfo
  {publisher} {Prentice Hall},\ \bibinfo {address} {Upper Saddle River, NJ},\
  \bibinfo {year} {2004})\BibitemShut {NoStop}%
\bibitem [{\citenamefont {Carr}\ \emph {et~al.}(2018)\citenamefont {Carr},
  \citenamefont {Massatt}, \citenamefont {Torrisi}, \citenamefont {Cazeaux},
  \citenamefont {Luskin},\ and\ \citenamefont {Kaxiras}}]{Carr2018Relaxation}%
  \BibitemOpen
  \bibfield  {author} {\bibinfo {author} {\bibfnamefont {S.}~\bibnamefont
  {Carr}}, \bibinfo {author} {\bibfnamefont {D.}~\bibnamefont {Massatt}},
  \bibinfo {author} {\bibfnamefont {S.~B.}\ \bibnamefont {Torrisi}}, \bibinfo
  {author} {\bibfnamefont {P.}~\bibnamefont {Cazeaux}}, \bibinfo {author}
  {\bibfnamefont {M.}~\bibnamefont {Luskin}}, \ and\ \bibinfo {author}
  {\bibfnamefont {E.}~\bibnamefont {Kaxiras}},\ }\href {\doibase
  10.1103/PhysRevB.98.224102} {\bibfield  {journal} {\bibinfo  {journal}
  {Physical Review B}\ }\textbf {\bibinfo {volume} {98}},\ \bibinfo {pages}
  {224102} (\bibinfo {year} {2018})}\BibitemShut {NoStop}%
\bibitem [{\citenamefont {Kohn}\ and\ \citenamefont
  {Sham}(1965)}]{kohn_self-consistent_1965}%
  \BibitemOpen
  \bibfield  {author} {\bibinfo {author} {\bibfnamefont {W.}~\bibnamefont
  {Kohn}}\ and\ \bibinfo {author} {\bibfnamefont {L.~J.}\ \bibnamefont
  {Sham}},\ }\href {\doibase 10.1103/PhysRev.140.A1133} {\bibfield  {journal}
  {\bibinfo  {journal} {Physical Review}\ }\textbf {\bibinfo {volume} {140}},\
  \bibinfo {pages} {A1133} (\bibinfo {year} {1965})}\BibitemShut {NoStop}%
\bibitem [{\citenamefont {Fang}\ \emph {et~al.}(2015)\citenamefont {Fang},
  \citenamefont {Kuate~Defo}, \citenamefont {Shirodkar}, \citenamefont {Lieu},
  \citenamefont {Tritsaris},\ and\ \citenamefont {Kaxiras}}]{Fang2015Ab}%
  \BibitemOpen
  \bibfield  {author} {\bibinfo {author} {\bibfnamefont {S.}~\bibnamefont
  {Fang}}, \bibinfo {author} {\bibfnamefont {R.}~\bibnamefont {Kuate~Defo}},
  \bibinfo {author} {\bibfnamefont {S.~N.}\ \bibnamefont {Shirodkar}}, \bibinfo
  {author} {\bibfnamefont {S.}~\bibnamefont {Lieu}}, \bibinfo {author}
  {\bibfnamefont {G.~A.}\ \bibnamefont {Tritsaris}}, \ and\ \bibinfo {author}
  {\bibfnamefont {E.}~\bibnamefont {Kaxiras}},\ }\href {\doibase
  10.1103/PhysRevB.92.205108} {\bibfield  {journal} {\bibinfo  {journal}
  {Physical Review B}\ }\textbf {\bibinfo {volume} {92}},\ \bibinfo {pages}
  {205108} (\bibinfo {year} {2015})}\BibitemShut {NoStop}%
\bibitem [{\citenamefont {Bradski}(2000)}]{Bradski2000OpenCV}%
  \BibitemOpen
  \bibfield  {author} {\bibinfo {author} {\bibfnamefont {G.}~\bibnamefont
  {Bradski}},\ }\href@noop {} {\bibfield  {journal} {\bibinfo  {journal} {Dr.
  Dobb's Journal of Software Tools}\ } (\bibinfo {year} {2000})}\BibitemShut
  {NoStop}%
\bibitem [{\citenamefont {Weininger}(1988)}]{Weininger1988SMILES}%
  \BibitemOpen
  \bibfield  {author} {\bibinfo {author} {\bibfnamefont {D.}~\bibnamefont
  {Weininger}},\ }\href {\doibase 10.1021/ci00057a005} {\bibfield  {journal}
  {\bibinfo  {journal} {Journal of Chemical Information and Computer Sciences}\
  }\textbf {\bibinfo {volume} {28}},\ \bibinfo {pages} {31} (\bibinfo {year}
  {1988})}\BibitemShut {NoStop}%
\bibitem [{\citenamefont {Systems}()}]{D.C.I.SystemsSMARTS}%
  \BibitemOpen
  \bibfield  {author} {\bibinfo {author} {\bibfnamefont {D.}~\bibnamefont
  {Systems}},\ }\href
  {https://www.daylight.com/dayhtml/doc/theory/theory.smarts.html} {\
  }\BibitemShut {NoStop}%
\bibitem [{\citenamefont {Tritsaris}\ \emph {et~al.}(2019)\citenamefont
  {Tritsaris}, \citenamefont {Xie}, \citenamefont {Rush}, \citenamefont {Carr},
  \citenamefont {Mattheakis},\ and\ \citenamefont
  {Kaxiras}}]{Tritsaris2019LAN}%
  \BibitemOpen
  \bibfield  {author} {\bibinfo {author} {\bibfnamefont {G.~A.}\ \bibnamefont
  {Tritsaris}}, \bibinfo {author} {\bibfnamefont {Y.}~\bibnamefont {Xie}},
  \bibinfo {author} {\bibfnamefont {A.~M.}\ \bibnamefont {Rush}}, \bibinfo
  {author} {\bibfnamefont {S.}~\bibnamefont {Carr}}, \bibinfo {author}
  {\bibfnamefont {M.}~\bibnamefont {Mattheakis}}, \ and\ \bibinfo {author}
  {\bibfnamefont {E.}~\bibnamefont {Kaxiras}},\ }\href
  {https://arxiv.org/abs/1910.03413v1} {\  (\bibinfo {year} {2019})},\ \bibinfo
  {note} {[Online; accessed 2019-10-08]}\BibitemShut {NoStop}%
\bibitem [{\citenamefont {Uchida}\ \emph {et~al.}(2014)\citenamefont {Uchida},
  \citenamefont {Furuya}, \citenamefont {Iwata},\ and\ \citenamefont
  {Oshiyama}}]{Uchida2014Atomic}%
  \BibitemOpen
  \bibfield  {author} {\bibinfo {author} {\bibfnamefont {K.}~\bibnamefont
  {Uchida}}, \bibinfo {author} {\bibfnamefont {S.}~\bibnamefont {Furuya}},
  \bibinfo {author} {\bibfnamefont {J.-I.}\ \bibnamefont {Iwata}}, \ and\
  \bibinfo {author} {\bibfnamefont {A.}~\bibnamefont {Oshiyama}},\ }\href
  {\doibase 10.1103/PhysRevB.90.155451} {\bibfield  {journal} {\bibinfo
  {journal} {Physical Review B}\ }\textbf {\bibinfo {volume} {90}} (\bibinfo
  {year} {2014}),\ 10.1103/PhysRevB.90.155451},\ \bibinfo {note} {[Online;
  accessed 2015-10-22]}\BibitemShut {NoStop}%
\bibitem [{\citenamefont {Nam}\ and\ \citenamefont
  {Koshino}(2017)}]{Nam2017Lattice}%
  \BibitemOpen
  \bibfield  {author} {\bibinfo {author} {\bibfnamefont {N.~N.~T.}\
  \bibnamefont {Nam}}\ and\ \bibinfo {author} {\bibfnamefont {M.}~\bibnamefont
  {Koshino}},\ }\href {\doibase 10.1103/PhysRevB.96.075311} {\bibfield
  {journal} {\bibinfo  {journal} {Physical Review B}\ }\textbf {\bibinfo
  {volume} {96}},\ \bibinfo {pages} {075311} (\bibinfo {year}
  {2017})}\BibitemShut {NoStop}%
\bibitem [{\citenamefont {Carr}\ \emph
  {et~al.}(2019{\natexlab{a}})\citenamefont {Carr}, \citenamefont {Fang},
  \citenamefont {Zhu},\ and\ \citenamefont {Kaxiras}}]{Carr2019Exact}%
  \BibitemOpen
  \bibfield  {author} {\bibinfo {author} {\bibfnamefont {S.}~\bibnamefont
  {Carr}}, \bibinfo {author} {\bibfnamefont {S.}~\bibnamefont {Fang}}, \bibinfo
  {author} {\bibfnamefont {Z.}~\bibnamefont {Zhu}}, \ and\ \bibinfo {author}
  {\bibfnamefont {E.}~\bibnamefont {Kaxiras}},\ }\href {\doibase
  10.1103/PhysRevResearch.1.013001} {\bibfield  {journal} {\bibinfo  {journal}
  {Physical Review Research}\ }\textbf {\bibinfo {volume} {1}},\ \bibinfo
  {pages} {013001} (\bibinfo {year} {2019}{\natexlab{a}})}\BibitemShut
  {NoStop}%
\bibitem [{\citenamefont {Carr}\ \emph
  {et~al.}(2019{\natexlab{b}})\citenamefont {Carr}, \citenamefont {Li},
  \citenamefont {Zhu}, \citenamefont {Kaxiras}, \citenamefont {Sachdev},\ and\
  \citenamefont {Kruchkov}}]{Carr2019Coexistence}%
  \BibitemOpen
  \bibfield  {author} {\bibinfo {author} {\bibfnamefont {S.}~\bibnamefont
  {Carr}}, \bibinfo {author} {\bibfnamefont {C.}~\bibnamefont {Li}}, \bibinfo
  {author} {\bibfnamefont {Z.}~\bibnamefont {Zhu}}, \bibinfo {author}
  {\bibfnamefont {E.}~\bibnamefont {Kaxiras}}, \bibinfo {author} {\bibfnamefont
  {S.}~\bibnamefont {Sachdev}}, \ and\ \bibinfo {author} {\bibfnamefont
  {A.}~\bibnamefont {Kruchkov}},\ }\href {http://arxiv.org/abs/1907.00952}
  {\bibfield  {journal} {\bibinfo  {journal} {arXiv:1907.00952 [cond-mat]}\ }
  (\bibinfo {year} {2019}{\natexlab{b}})},\ \bibinfo {note} {arXiv:
  1907.00952}\BibitemShut {NoStop}%
\bibitem [{\citenamefont {Khalaf}\ \emph {et~al.}(2019)\citenamefont {Khalaf},
  \citenamefont {Kruchkov}, \citenamefont {Tarnopolsky},\ and\ \citenamefont
  {Vishwanath}}]{Khalaf2019Magic}%
  \BibitemOpen
  \bibfield  {author} {\bibinfo {author} {\bibfnamefont {E.}~\bibnamefont
  {Khalaf}}, \bibinfo {author} {\bibfnamefont {A.~J.}\ \bibnamefont
  {Kruchkov}}, \bibinfo {author} {\bibfnamefont {G.}~\bibnamefont
  {Tarnopolsky}}, \ and\ \bibinfo {author} {\bibfnamefont {A.}~\bibnamefont
  {Vishwanath}},\ }\href {\doibase 10.1103/PhysRevB.100.085109} {\bibfield
  {journal} {\bibinfo  {journal} {Physical Review B}\ }\textbf {\bibinfo
  {volume} {100}},\ \bibinfo {pages} {085109} (\bibinfo {year}
  {2019})}\BibitemShut {NoStop}%
\bibitem [{\citenamefont {Cea}\ \emph {et~al.}(2019)\citenamefont {Cea},
  \citenamefont {Walet},\ and\ \citenamefont {Guinea}}]{Cea2019Twists}%
  \BibitemOpen
  \bibfield  {author} {\bibinfo {author} {\bibfnamefont {T.}~\bibnamefont
  {Cea}}, \bibinfo {author} {\bibfnamefont {N.~R.}\ \bibnamefont {Walet}}, \
  and\ \bibinfo {author} {\bibfnamefont {F.}~\bibnamefont {Guinea}},\ }\href
  {\doibase 10.1021/acs.nanolett.9b03335} {\bibfield  {journal} {\bibinfo
  {journal} {Nano Letters}\ }\textbf {\bibinfo {volume} {19}},\ \bibinfo
  {pages} {8683} (\bibinfo {year} {2019})}\BibitemShut {NoStop}%
\bibitem [{\citenamefont {Lee}\ \emph {et~al.}(2019)\citenamefont {Lee},
  \citenamefont {Khalaf}, \citenamefont {Liu}, \citenamefont {Liu},
  \citenamefont {Hao}, \citenamefont {Kim},\ and\ \citenamefont
  {Vishwanath}}]{Lee2019Theory}%
  \BibitemOpen
  \bibfield  {author} {\bibinfo {author} {\bibfnamefont {J.~Y.}\ \bibnamefont
  {Lee}}, \bibinfo {author} {\bibfnamefont {E.}~\bibnamefont {Khalaf}},
  \bibinfo {author} {\bibfnamefont {S.}~\bibnamefont {Liu}}, \bibinfo {author}
  {\bibfnamefont {X.}~\bibnamefont {Liu}}, \bibinfo {author} {\bibfnamefont
  {Z.}~\bibnamefont {Hao}}, \bibinfo {author} {\bibfnamefont {P.}~\bibnamefont
  {Kim}}, \ and\ \bibinfo {author} {\bibfnamefont {A.}~\bibnamefont
  {Vishwanath}},\ }\href {https://www.nature.com/articles/s41467-019-12981-1}
  {\bibfield  {journal} {\bibinfo  {journal} {Nature Communications}\ }\textbf
  {\bibinfo {volume} {10}} (\bibinfo {year} {2019})}\BibitemShut {NoStop}%
\bibitem [{\citenamefont {Shen}\ \emph {et~al.}(2019)\citenamefont {Shen},
  \citenamefont {Li}, \citenamefont {Wang}, \citenamefont {Zhao}, \citenamefont
  {Tang}, \citenamefont {Liu}, \citenamefont {Tian}, \citenamefont {Chu},
  \citenamefont {Watanabe}, \citenamefont {Taniguchi}, \citenamefont {Yang},
  \citenamefont {Meng}, \citenamefont {Shi},\ and\ \citenamefont
  {Zhang}}]{Shen2019Observation}%
  \BibitemOpen
  \bibfield  {author} {\bibinfo {author} {\bibfnamefont {C.}~\bibnamefont
  {Shen}}, \bibinfo {author} {\bibfnamefont {N.}~\bibnamefont {Li}}, \bibinfo
  {author} {\bibfnamefont {S.}~\bibnamefont {Wang}}, \bibinfo {author}
  {\bibfnamefont {Y.}~\bibnamefont {Zhao}}, \bibinfo {author} {\bibfnamefont
  {J.}~\bibnamefont {Tang}}, \bibinfo {author} {\bibfnamefont {J.}~\bibnamefont
  {Liu}}, \bibinfo {author} {\bibfnamefont {J.}~\bibnamefont {Tian}}, \bibinfo
  {author} {\bibfnamefont {Y.}~\bibnamefont {Chu}}, \bibinfo {author}
  {\bibfnamefont {K.}~\bibnamefont {Watanabe}}, \bibinfo {author}
  {\bibfnamefont {T.}~\bibnamefont {Taniguchi}}, \bibinfo {author}
  {\bibfnamefont {R.}~\bibnamefont {Yang}}, \bibinfo {author} {\bibfnamefont
  {Z.~Y.}\ \bibnamefont {Meng}}, \bibinfo {author} {\bibfnamefont
  {D.}~\bibnamefont {Shi}}, \ and\ \bibinfo {author} {\bibfnamefont
  {G.}~\bibnamefont {Zhang}},\ }\href {http://arxiv.org/abs/1903.06952}
  {\bibfield  {journal} {\bibinfo  {journal} {arXiv:1903.06952 [cond-mat]}\ }
  (\bibinfo {year} {2019})},\ \bibinfo {note} {arXiv: 1903.06952}\BibitemShut
  {NoStop}%
\bibitem [{\citenamefont {Cao}\ \emph {et~al.}(2019)\citenamefont {Cao},
  \citenamefont {Rodan-Legrain}, \citenamefont {Rubies-Bigorda}, \citenamefont
  {Park}, \citenamefont {Watanabe}, \citenamefont {Taniguchi},\ and\
  \citenamefont {Jarillo-Herrero}}]{Cao2019Electric}%
  \BibitemOpen
  \bibfield  {author} {\bibinfo {author} {\bibfnamefont {Y.}~\bibnamefont
  {Cao}}, \bibinfo {author} {\bibfnamefont {D.}~\bibnamefont {Rodan-Legrain}},
  \bibinfo {author} {\bibfnamefont {O.}~\bibnamefont {Rubies-Bigorda}},
  \bibinfo {author} {\bibfnamefont {J.~M.}\ \bibnamefont {Park}}, \bibinfo
  {author} {\bibfnamefont {K.}~\bibnamefont {Watanabe}}, \bibinfo {author}
  {\bibfnamefont {T.}~\bibnamefont {Taniguchi}}, \ and\ \bibinfo {author}
  {\bibfnamefont {P.}~\bibnamefont {Jarillo-Herrero}},\ }\href
  {http://arxiv.org/abs/1903.08596} {\bibfield  {journal} {\bibinfo  {journal}
  {arXiv:1903.08596 [cond-mat]}\ } (\bibinfo {year} {2019})},\ \bibinfo {note}
  {arXiv: 1903.08596}\BibitemShut {NoStop}%
\bibitem [{\citenamefont {Pedregosa}\ \emph {et~al.}(2011)\citenamefont
  {Pedregosa}, \citenamefont {Varoquaux}, \citenamefont {Gramfort},
  \citenamefont {Michel}, \citenamefont {Thirion}, \citenamefont {Grisel},
  \citenamefont {Blondel}, \citenamefont {Prettenhofer}, \citenamefont {Weiss},
  \citenamefont {Dubourg}, \citenamefont {Vanderplas}, \citenamefont {Passos},
  \citenamefont {Cournapeau}, \citenamefont {Brucher}, \citenamefont {Perrot},\
  and\ \citenamefont {Duchesnay}}]{Pedregosa2011Scikit-learn:}%
  \BibitemOpen
  \bibfield  {author} {\bibinfo {author} {\bibfnamefont {F.}~\bibnamefont
  {Pedregosa}}, \bibinfo {author} {\bibfnamefont {G.}~\bibnamefont
  {Varoquaux}}, \bibinfo {author} {\bibfnamefont {A.}~\bibnamefont {Gramfort}},
  \bibinfo {author} {\bibfnamefont {V.}~\bibnamefont {Michel}}, \bibinfo
  {author} {\bibfnamefont {B.}~\bibnamefont {Thirion}}, \bibinfo {author}
  {\bibfnamefont {O.}~\bibnamefont {Grisel}}, \bibinfo {author} {\bibfnamefont
  {M.}~\bibnamefont {Blondel}}, \bibinfo {author} {\bibfnamefont
  {P.}~\bibnamefont {Prettenhofer}}, \bibinfo {author} {\bibfnamefont
  {R.}~\bibnamefont {Weiss}}, \bibinfo {author} {\bibfnamefont
  {V.}~\bibnamefont {Dubourg}}, \bibinfo {author} {\bibfnamefont
  {J.}~\bibnamefont {Vanderplas}}, \bibinfo {author} {\bibfnamefont
  {A.}~\bibnamefont {Passos}}, \bibinfo {author} {\bibfnamefont
  {D.}~\bibnamefont {Cournapeau}}, \bibinfo {author} {\bibfnamefont
  {M.}~\bibnamefont {Brucher}}, \bibinfo {author} {\bibfnamefont
  {M.}~\bibnamefont {Perrot}}, \ and\ \bibinfo {author} {\bibfnamefont
  {{\'E}.}~\bibnamefont {Duchesnay}},\ }\href@noop {} {\bibfield  {journal}
  {\bibinfo  {journal} {Journal of Machine Learning Research}\ }\textbf
  {\bibinfo {volume} {12}},\ \bibinfo {pages} {2825} (\bibinfo {year}
  {2011})}\BibitemShut {NoStop}%
\bibitem [{\citenamefont {Elton}\ \emph {et~al.}(2018)\citenamefont {Elton},
  \citenamefont {Boukouvalas}, \citenamefont {Butrico}, \citenamefont {Fuge},\
  and\ \citenamefont {Chung}}]{Elton2018Applying}%
  \BibitemOpen
  \bibfield  {author} {\bibinfo {author} {\bibfnamefont {D.~C.}\ \bibnamefont
  {Elton}}, \bibinfo {author} {\bibfnamefont {Z.}~\bibnamefont {Boukouvalas}},
  \bibinfo {author} {\bibfnamefont {M.~S.}\ \bibnamefont {Butrico}}, \bibinfo
  {author} {\bibfnamefont {M.~D.}\ \bibnamefont {Fuge}}, \ and\ \bibinfo
  {author} {\bibfnamefont {P.~W.}\ \bibnamefont {Chung}},\ }\href {\doibase
  10.1038/s41598-018-27344-x} {\bibfield  {journal} {\bibinfo  {journal}
  {Scientific Reports}\ }\textbf {\bibinfo {volume} {8}},\ \bibinfo {pages} {1}
  (\bibinfo {year} {2018})}\BibitemShut {NoStop}%
\bibitem [{\citenamefont {Roch}\ \emph {et~al.}(2018)\citenamefont {Roch},
  \citenamefont {Häse}, \citenamefont {Kreisbeck}, \citenamefont
  {Tamayo-Mendoza}, \citenamefont {Yunker}, \citenamefont {Hein},\ and\
  \citenamefont {Aspuru-Guzik}}]{roch_chemos:_2018}%
  \BibitemOpen
  \bibfield  {author} {\bibinfo {author} {\bibfnamefont {L.~M.}\ \bibnamefont
  {Roch}}, \bibinfo {author} {\bibfnamefont {F.}~\bibnamefont {Häse}},
  \bibinfo {author} {\bibfnamefont {C.}~\bibnamefont {Kreisbeck}}, \bibinfo
  {author} {\bibfnamefont {T.}~\bibnamefont {Tamayo-Mendoza}}, \bibinfo
  {author} {\bibfnamefont {L.~P.~E.}\ \bibnamefont {Yunker}}, \bibinfo {author}
  {\bibfnamefont {J.~E.}\ \bibnamefont {Hein}}, \ and\ \bibinfo {author}
  {\bibfnamefont {A.}~\bibnamefont {Aspuru-Guzik}},\ }\href {\doibase
  10.26434/chemrxiv.5953606.v1} {\  (\bibinfo {year} {2018}),\
  10.26434/chemrxiv.5953606.v1}\BibitemShut {NoStop}%
\bibitem [{\citenamefont {Dunn}\ \emph {et~al.}(2019)\citenamefont {Dunn},
  \citenamefont {Brenneck},\ and\ \citenamefont {Jain}}]{Dunn_2019}%
  \BibitemOpen
  \bibfield  {author} {\bibinfo {author} {\bibfnamefont {A.}~\bibnamefont
  {Dunn}}, \bibinfo {author} {\bibfnamefont {J.}~\bibnamefont {Brenneck}}, \
  and\ \bibinfo {author} {\bibfnamefont {A.}~\bibnamefont {Jain}},\ }\href
  {\doibase 10.1088/2515-7639/ab0c3d} {\bibfield  {journal} {\bibinfo
  {journal} {Journal of Physics: Materials}\ }\textbf {\bibinfo {volume} {2}},\
  \bibinfo {pages} {034002} (\bibinfo {year} {2019})}\BibitemShut {NoStop}%
\bibitem [{\citenamefont {Geurts}\ \emph {et~al.}(2006)\citenamefont {Geurts},
  \citenamefont {Ernst},\ and\ \citenamefont {Wehenkel}}]{Geurts2006Extremely}%
  \BibitemOpen
  \bibfield  {author} {\bibinfo {author} {\bibfnamefont {P.}~\bibnamefont
  {Geurts}}, \bibinfo {author} {\bibfnamefont {D.}~\bibnamefont {Ernst}}, \
  and\ \bibinfo {author} {\bibfnamefont {L.}~\bibnamefont {Wehenkel}},\ }\href
  {\doibase 10.1007/s10994-006-6226-1} {\bibfield  {journal} {\bibinfo
  {journal} {Machine Learning}\ }\textbf {\bibinfo {volume} {63}},\ \bibinfo
  {pages} {3} (\bibinfo {year} {2006})}\BibitemShut {NoStop}%
\bibitem [{\citenamefont {Zhu}\ \emph {et~al.}(2019)\citenamefont {Zhu},
  \citenamefont {Cazeaux}, \citenamefont {Luskin},\ and\ \citenamefont
  {Kaxiras}}]{Zhu2019Moire}%
  \BibitemOpen
  \bibfield  {author} {\bibinfo {author} {\bibfnamefont {Z.}~\bibnamefont
  {Zhu}}, \bibinfo {author} {\bibfnamefont {P.}~\bibnamefont {Cazeaux}},
  \bibinfo {author} {\bibfnamefont {M.}~\bibnamefont {Luskin}}, \ and\ \bibinfo
  {author} {\bibfnamefont {E.}~\bibnamefont {Kaxiras}},\ }\href
  {http://arxiv.org/abs/1911.05324} {\bibfield  {journal} {\bibinfo  {journal}
  {arXiv:1911.05324 [cond-mat]}\ } (\bibinfo {year} {2019})},\ \bibinfo {note}
  {arXiv: 1911.05324}\BibitemShut {NoStop}%
\end{thebibliography}%

\end{document}